%% file: main.tex
\documentclass[modern]{aastex631}

\input{defs.tex}

\shorttitle{Narrow-line AGN}
\shortauthors{Blanton et al.}

\graphicspath{{./}}

\begin{document}

\title{The Eddington Ratio Distribution of Narrow-line Active Galactic
  Nuclei}

\author[0000-0003-1641-6222]{Michael R.~Blanton}
\affiliation{
The Observatories of the Carnegie Institution for Science, 813 Santa Barbara Street, \\
Pasadena, CA, 91101, USA}
\affiliation{Center for Cosmology and Particle Physics,
Department of Physics,
New York University, \\
726 Broadway Rm. 1005,
New York, NY 10003, USA}

\author[0009-0008-7407-0665]{Arjun Suresh}
\affiliation{Center for Cosmology and Particle Physics,
Department of Physics,
New York University, \\
726 Broadway Rm. 1005,
New York, NY 10003, USA}

\author[0000-0003-1809-6920]{Kyle B. Westfall}
\affiliation{University of California Observatories, University of California, Santa Cruz, 1156 High St., \\
Santa Cruz, CA 95064, USA}

\author[0000-0002-4069-6415]{Dou Liu}
\affiliation{Wells Fargo, 550 S Tryon St, Charlotte, NC 28202, USA}

\author[0000-0002-2733-4559]{John Moustakas}
\affiliation{Department of Physics and Astronomy, Siena College, 515
  Loudon Road, \\
  Loudonville, NY 12211, USA}

\begin{abstract}
We measure the Eddington ratio distribution of local optical
narrow-line active galactic nuclei (AGN) as a function of host galaxy
properties, as a potential test of galaxy formation theories of AGN
feedback. We extract central emission-line fluxes using data from the
Mapping Nearby Galaxies at APO (MaNGA) sample of the Sloan Digital Sky
Survey IV Data Release 17. Using the line ratio diagnostic techniques
of \citet{ji20a}, we identify AGN galaxies and determine their
\hb\ and \oiii\ line luminosities. For all galaxies not identified as
AGN, we determine the threshold line luminosity they would have needed
to be identified as AGN. These luminosity thresholds allow us to
account for selection effects that otherwise would lead to strongly
biased results.  From the \hb\ luminosities and luminosity detection
thresholds, accounting for selection effects, we measure the
luminosity and Eddington ratio distributions of Seyferts as a function
of specific star formation rate (sSFR) and stellar mass. Defining
$F_{\rm AGN}$ as the occurrence rate of AGN above a fixed Eddington
ratio of $10^{-3}$, we find that $F_{\rm AGN}$ is constant or
increasing with stellar mass for star forming galaxies and declines
strongly with stellar mass for quiescent galaxies. At stellar masses
$\log_{10} M_\ast > 10.25$, the occurrence rate increases
monotonically with sSFR. At low statistical significance, in our
lowest mass bins $9.25 < \log_{10} M_\ast < 10.25$, $F_{\rm AGN}$
peaks at intermediate sSFR. These patterns reveal a complicated
dependence of AGN activity on galaxy properties for theoretical models
to explain.
\end{abstract}

\section{Introduction} \label{sec:intro}

At the centers of essentially all massive galaxies are supermassive
black holes, with masses that can exceed $10^9$ $M_\odot$. During the 
growth of these black holes, they can emit jets, winds, and radiation, 
sometimes outshining the galaxy itself, a condition known as an active 
galactic nucleus (AGN).

Modern models of galaxy formation rely on this emission to explain
basic phenomena of galaxies, like their distribution of luminosities,
colors, and, star-formation rates (SFRs; for reviews see
\citealt{fabian12, somerville15a}). Essentially, this AGN feedback is
thought to modulate and in some cases shut down star formation.
Therefore the relationship between the AGN activity and the other
properties of the galaxies is a potential observable that could
distinguish competing models of galaxy formation.

Virtually all galaxies have supermassive black holes, but only a small
fraction have detectable AGN at any given time, even at higher
redshifts. The general interpretation of this difference is that
quasars have limited periods of activity, so that at any given time
only a fraction of galaxies have luminous AGN (\citealt{yu02a,
  shankar09a, hickox14a}).  The fraction of the time a galaxy spends
as an AGN is referred to as the ``duty cycle."  The total amount of
time a galaxy spends as an AGN is often called the ``lifetime," though
sometimes this term is reserved for the length of individual episodes
(e.g. \citealt{khrykin21a}).  The numerical values of these quantities
depend on the definition of what an AGN is---e.g. what the minimum
luminosity, Eddington ratio, or other characteristic of the emission
has to be for us to identify it as an AGN.

One quantification of AGN demographics is the distribution of their
luminosities and/or Eddington ratios. In the ansatz we adopt in this
paper, every galaxy is an AGN of some luminosity, with a broad
distribution of luminosities, but for most galaxies most of the time
the AGN luminosity is undetectably low. The occurrence rate of AGN
above some threshold of activity is then interpretable as the mean
duty cycle of AGN above that threshold. If we want to understand how
AGN and galaxy relate to each other and assess theoretical models of
galaxy formation, key measurements are how these occurrence rates vary
with galaxy mass, SFR, and other properties.

This work is part of a larger program to reassess the demographics of
AGN with respect to their host galaxies, which began with
\citet{suresh24a}.  There are many manifestations of AGN, as X-ray,
UV, optical broad- and narrow-line, mid-infrared UV, and radio sources,
which need to be considered.  Furthermore, patterns of AGN activity
are known to vary strongly with redshift, peaking at redshifts $z\sim
2$--$3$. An enormous amount of work has gone into selecting samples of
AGN and observing them out to high redshift. However, because the
selection process is complex, it is difficult to infer relationships
between the AGN samples and their host galaxies.

\citet{aird12a} argue convincingly that selecting samples by their AGN
luminosity, as many X-ray, radio, and optical studies do, means that
we are catching only galaxies in the rarer moments when they are most
strongly active, and makes it difficult to understand the relationship
between AGN and the galaxy host population. \citet{hickox14a}
explicitly demonstrate the resulting importance of selecting AGN from
statistically complete galaxy samples and measuring the distribution
of AGN properties conditioned on galaxy properties, rather than the
converse.  Among other problems, it is difficult to constrain
quantities like the duty cycle from samples selected by AGN
luminosity: because only luminous AGN are in the sample, the luminous
AGN fraction cannot be directly determined relative to the full
distribution of galaxies. In such samples, it is also difficult to
account for selection effects associated with finding lower-luminosity
AGN that might be outshone by their host galaxy. For all these
reasons, our approach is to start with a galaxy sample and measure the
Eddington ratio distribution of the AGN, accounting for the selection
effects.

In this paper we concern ourselves with a single AGN manifestation,
optical narrow-line emission. We use a low-redshift ($z\sim 0.03$)
sample that will allow us to have as much understanding of selection
effects as we can while still retaining a statistically useful sample.
The primary point of this paper is to accurately measure the
\oiiilam\ and \hb\ narrow-line luminosity distributions and the
corresponding Eddington ratio distributions as a function of galaxy
stellar mass and SFR. Our methodology follows 
closely that of \citet{trump15a}, who performed a similar analysis on
the Sloan Digital Sky Survey Legacy Survey
(\citealt{york00a}). \citet{suresh24a} have conducted a similar study
using the 1.4 GHz radio emission of AGN.

We make use of the Mapping Nearby Galaxies at APO (MaNGA;
\citealt{bundy15a}) program, part of the Sloan Digital Sky Survey IV
(SDSS-IV; \citealt{blanton17a}). The sample we use  is much smaller 
than  the SDSS Legacy Survey Main Sample (\citealt{strauss02a}) or the many
millions in the Dark Energy Spectroscopic Instrument (DESI;
\citealt{abareshi22a}) Bright-time Galaxy Survey (BGS)---about $10^4$ galaxies 
in MaNGA versus $10^6$ in SDSS Legacy and $10^7$ in DESI BGS. But MaNGA has far
better measurements for each galaxy---so if there is any sample for
which we can disentangle the AGN selection effects, it is this one.

In Section \ref{sec:data}, we describe our MaNGA central spectrum line
measurements. In Section \ref{sec:agn}, we discuss our narrow-line AGN
selection, how we determine detection thresholds, and how we measure
Eddington ratios. In Section \ref{sec:luminosities}, we present the
luminosities, Eddington ratios, and AGN detection thresholds for both
quantities for the MaNGA sample. In Section \ref{sec:ldist}, we
estimate the intrinsic narrow-line luminosity distributions inferred
from the sample. In Section \ref{sec:edist}, we do the same for
the Eddington ratio distributions. We compare to previous results in
Section \ref{sec:comparison} and present our conclusions in Section
\ref{sec:conclusion}.

The cosmological parameters we use to calculate luminosities are those
of \citet{Planck18.VI} (far right column of their Table 2), as
implemented by the software in {\tt astropy.cosmology.Planck18}.

\section{MaNGA Data}
\label{sec:data}

\subsection{MaNGA Overview}
\label{sec:data-overview}

MaNGA is an integral field survey that observed around $10^4$ galaxies
during dark time at the Sloan Foundation Telescope at Apache Point
Observatory, New Mexico, as part of SDSS-IV (2014--2020). It used 17
optical fiber bundles of varying sizes (19, 37, 61, 91, and 127
fibers) plugged into spectroscopic plug-plates and feeding the Baryon
Oscillation Spectroscopic Survey (BOSS) spectrographs
(\citealt{smee13a, drory15a}).

The sample spans stellar masses of about $10^9$ to $10^{11}$
$M_\odot$, and includes all galaxy types in that range in its Primary,
Secondary, and Color-Enhanced samples. The target selection is not
uniform for a volume-limited sample; \citet{wake17a} provide a
description of the selection and the catalogs.


Using a suite of dithered observations, the Data Reduction Pipeline
extracted each fiber spectrum and built a data cube for each galaxy
(\citealt{law15a, law16a}). A Data Analysis Pipeline (DAP) measured
stellar and gas velocity and velocity dispersion maps, as well as
spectral index and emission-line flux maps (\citealt{belfiore19a,
  westfall19a}).

\citet{sanchez22a} ran their Pipe3D software on the data cubes to
create a value-added product containing alternative maps and
high-level derived parameters. These parameters include stellar
masses, SFRs based on \ha\ emission, and SFRs based on stellar
population synthesis models fit to the stellar continuum. In this
paper we use the Pipe3D SFR, specifically its H$\alpha$-based
estimate, excluding spaxels with strong emission-line ratios
consistent with Seyfert ionization (i.e. their {\tt log\_SFR\_SF}).

For this work we applied the data cube reconstruction methods of
\citet{liu20a}, which achieve slightly higher resolution than those
found in DR17 and yield zero off-diagonal covariances between flux
values at a given wavelength. We ran the DAP on these cubes, using the
same version as used for DR17 with some minor changes ({\tt
  3.1.0}). The most significant change to the DAP is that we set the
off-diagonal covariances to zero in the calculation, which is accurate
for our new cubes. We also added a check to mask a very small number
of very high inverse-variance pixels that appeared owing to an error
in our cube reconstruction. We have evaluated the results of this
paper using both our measurements and the DR17 measurements, and we do
not find any significant differences.

\subsection{Central Emission-line Fluxes}
\label{sec:data-fluxes}

To find AGN we search for central ionized gas emission with line
ratios consistent with AGN ionization spectra. The lines we consider
here are H$\alpha$, H$\beta$, \oiiilam, \niilam, and \siidoublet.

We measure central line fluxes from the DAP maps, using a post-spread
function (PSF)-weighted aperture:
\begin{equation}
f_{c} = \frac{\sum_i f_i {\rm PSF}_i}{\sum_i {\rm PSF}_i^2}
\end{equation}
where $i$ indexes the pixels.  We limit the PSF contributions to
within 3.5$''$ radius of the center, at which point the weighting
is small (of order 0.05).

We calculate the uncertainties from the variances reported by the DAP
for the individual map pixels (\citealt{belfiore19a}, Section 3) added
in quadrature using the weights. In calculating uncertainties in the
line ratios, we use simple error propagation based on small errors.

We consider a measurement ``good'' if three conditions are met:
if the redshift $z>0$ for the galaxy; if, for each of the six lines
mentioned above, there are three or fewer bad pixels contributing to the
central line flux; and if the flux has a positive inverse variance (even 
if the flux is consistent with zero). Otherwise we do not analyze the galaxy. 
There are 9,971 MaNGA observations that pass this criterion.  For 
the purposes of the AGN identification below, we consider a line 
``detected'' if its flux is measured at 2$\sigma$ or higher significance.

\subsection{MaNGA Sample}
\label{sec:manga-sample}

For our determination of the distribution of luminosities and
Eddington ratios, we use only MaNGA galaxies in its main statistical
samples. There are other targets observed by MaNGA, but using these
could lead to a statistical bias, since they are selected on known
galaxy properties, including in some cases that they are known to be
AGN.

Specifically, we use galaxies in the Primary, Secondary, or
Color-Enhanced samples as described by \citet{wake17a}. We use the
subsampled Secondary sample, which we identify as galaxies with a
selection weight {\tt ESRWEIGHT} $> 0$; this cut eliminates 515
galaxies. The final sample contains 9,029 galaxies. When we need black
hole masses we further cut the sample to stellar velocity
dispersions $\sigma_e>60$ km~s$^{-1}$, as measured by DAP within 1
$R_e$, which leaves 5,678 galaxies.

\begin{figure}
\begin{center}
\includegraphics[width=0.98\textwidth]{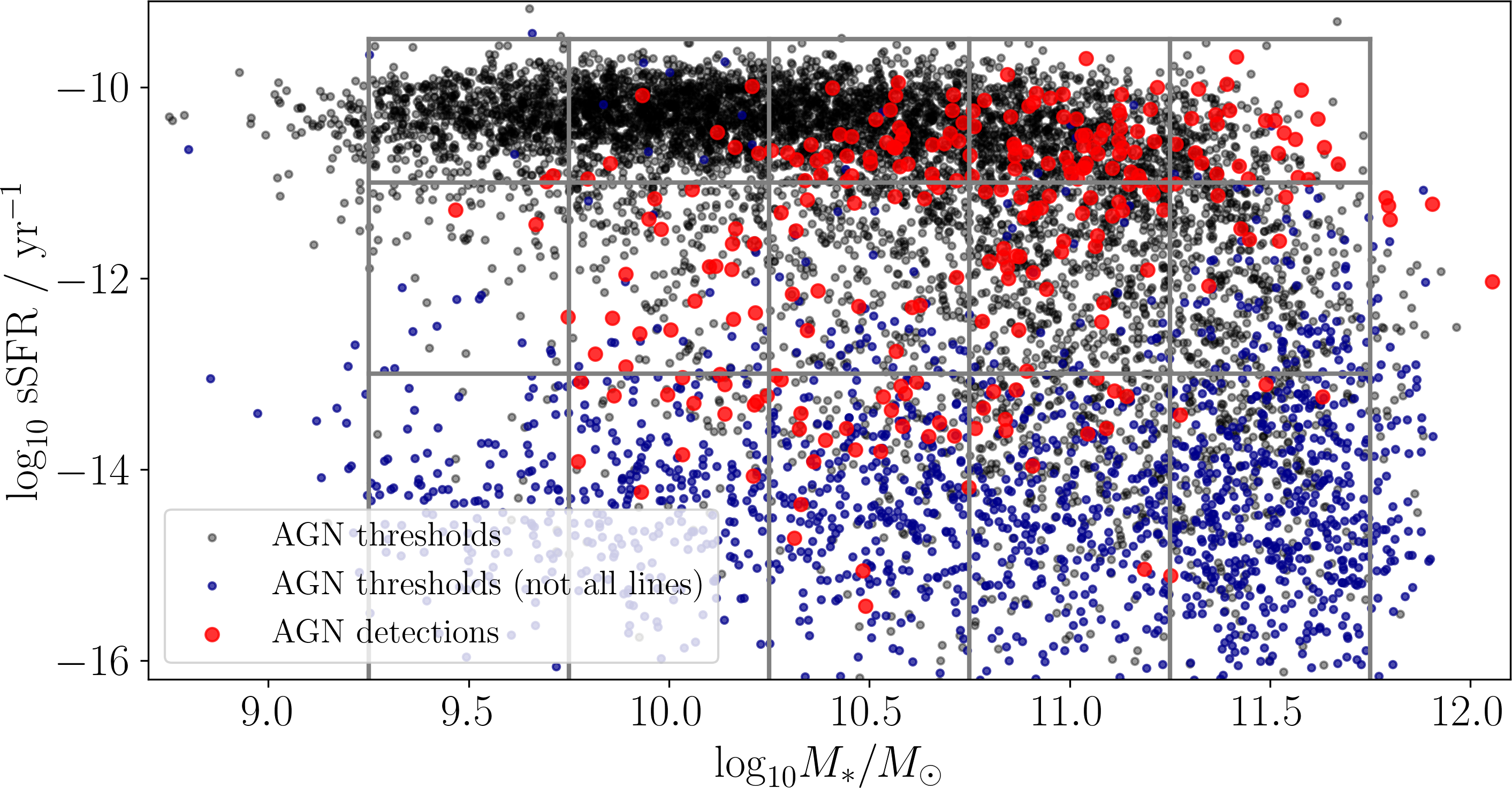}
\end{center}
\caption{\label{fig:mssfr} Stellar masses and sSFRs for the MaNGA
  sample used here. Red symbols show galaxies with AGN detections,
  black symbols show galaxies with AGN detection thresholds 
  (which have all necessary line ratios detected), and blue 
  symbols show  galaxies with AGN detection thresholds 
  (without all necessary line ratios detected).
  The boxes
  show divisions between stellar mass and sSFR samples we use to study
  the AGN luminosity and Eddington ratio distribution dependence on
  galaxy properties.}
\end{figure}

Figure \ref{fig:mssfr} shows the Pipe3D stellar masses and specific
SFRs (sSFRs) for the sample of 9,029 galaxies. The red symbols show
the galaxies with AGN detections (see details on the selection below),
and the black symbols show the galaxies with AGN detection thresholds.

\section{Identifying Narrow-line AGN}
\label{sec:agn}

\subsection{Line Ratio Diagnostics}
\label{sec:agn-line-ratio}

To identify AGN, we use the line ratio diagnostics defined by
\citet{ji20a}.  These diagnostics involve the same line ratios used by
\citet{baldwin81a} and \citet{veilleux87a}, namely \oiiihb\ (using
\oiiilam), \niiha\ (using \niilam), and \siiha\ (using
\siidoublet). These line ratios are each relatively insensitive to
spectrophotometric calibration errors or dust. The line ratio diagrams
of \citet{baldwin81a} and \citet{veilleux87a} typically are used as
two separate two-dimensional classification tools. In contrast,
\citet{ji20a} use three line ratios as a single three-dimensional
classification space.

Using the photoionization model \textsc{CLOUDY} \citet{ji20a} found an
informative reprojection of the three-dimensional space of these line
ratios. We use this reprojection $P1$, $P2$, and $P3$, which allows a
clean classification of spectra into those with ionizing spectra
consistent with star formation, LINER, and Seyfert sources. It is
defined as
\begin{eqnarray}
P1 &=& 0.63 \log_{10}\left(\frac{\nii}{\ha}\right) 
 + 0.51 \log_{10}\left(\frac{\sii}{\ha}\right)
 + 0.59 \log_{10}\left(\frac{\oiii}{\hb}\right) \cr
P2 &=& - 0.63 \log_{10}\left(\frac{\nii}{\ha}\right) 
 + 0.78 \log_{10}\left(\frac{\sii}{\ha}\right) \cr
P3 &=& - 0.46 \log_{10}\left(\frac{\nii}{\ha}\right) 
 - 0.37 \log_{10}\left(\frac{\sii}{\ha}\right)
 + 0.81 \log_{10}\left(\frac{\oiii}{\hb}\right)
\end{eqnarray}
\citet{ji20a} analyze the interpretation of this reprojection
thoroughly in terms of ionization models using both
star-formation-related ionization spectra (i.e. stellar populations)
and Seyfert/LINER-related ionization spectra (power-law spectra). A
simplified version of their interpretation is as follows. $P1$
correlates with the hardness of the ionization spectrum, and separates
star-formation-related ionization spectra (low $P1$) from LINERs and
Seyferts (high $P1$). $P2$ is negatively correlated with metallicity
for all classes. $P3$ is correlated well with the ionization parameter
$U=\Phi_{i} / n_H c$, where $\Phi_i$ is the flux of ionizing photons
per unit area per unit time, $n_H$ is the hydrogen atom density, and
$c$ is the speed of light. For star-formation-related ionization (low
$P1$), $P3$ also anticorrelates with metallicity in the models of
\citet{ji20a}, because metallicity determines the ionizing
spectrum. For the LINER and Seyfert models (high $P1$), $P3$
correlates primarily with $U$.

Thus, we can select Seyfert galaxies as high-$P1$, high-$P3$ galaxies;
in \citet{ji20a}, the left panel of their Figure 15 illustrates the
distribution for SDSS Legacy Main Sample galaxies
(\citealt{strauss02a}).  In this paper, we consider $P1 > -0.3$ and
$P3 > 0.55$ to define a Seyfert. We assume that the lower-$P3$,
LINER-like sources are powered by sources other than AGN (see
\citealt{sarzi10a, yan12a, belfiore16a} and references therein).

\begin{figure}
\begin{center}
\includegraphics[width=0.49\textwidth]{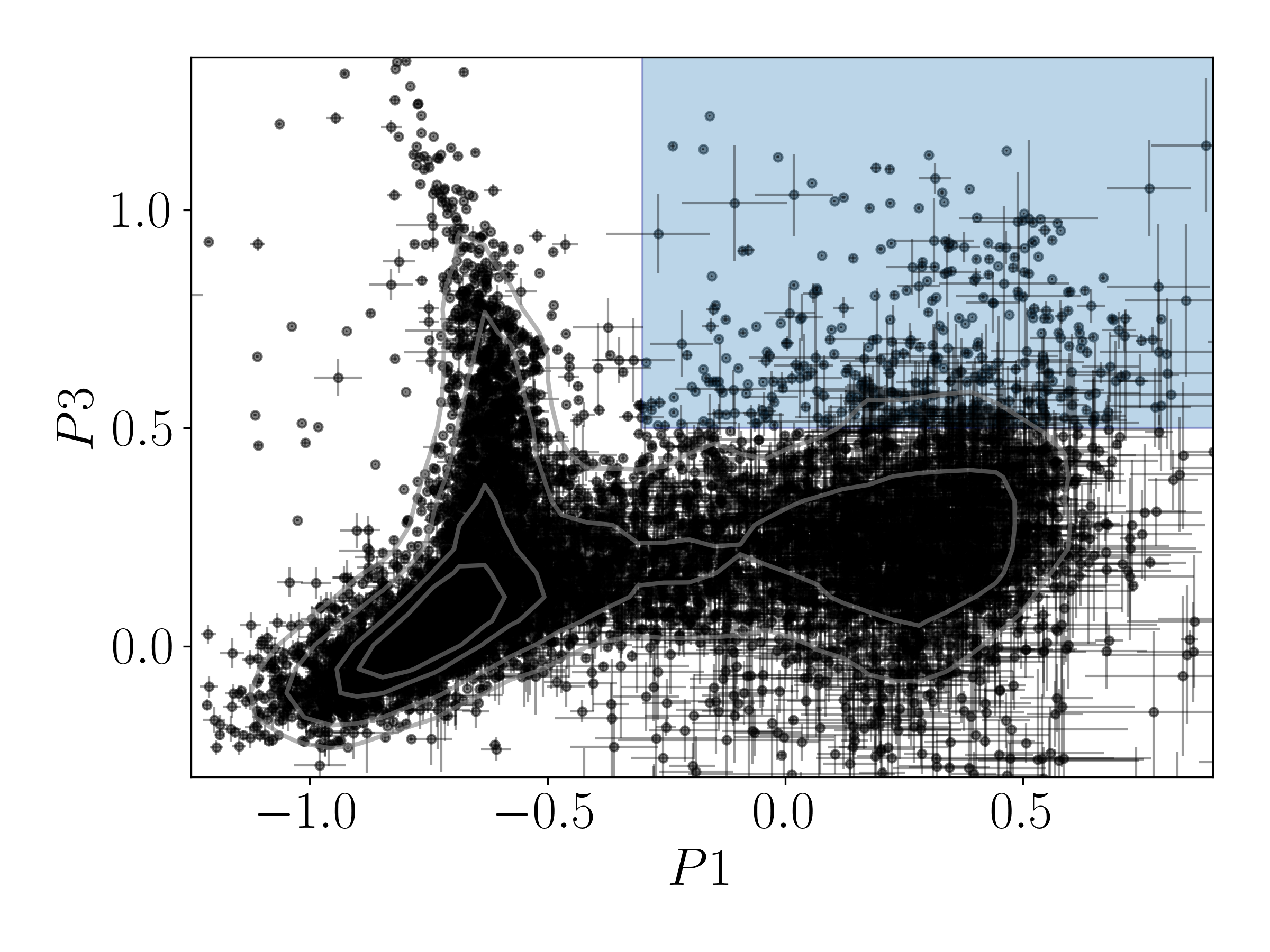}
\includegraphics[width=0.49\textwidth]{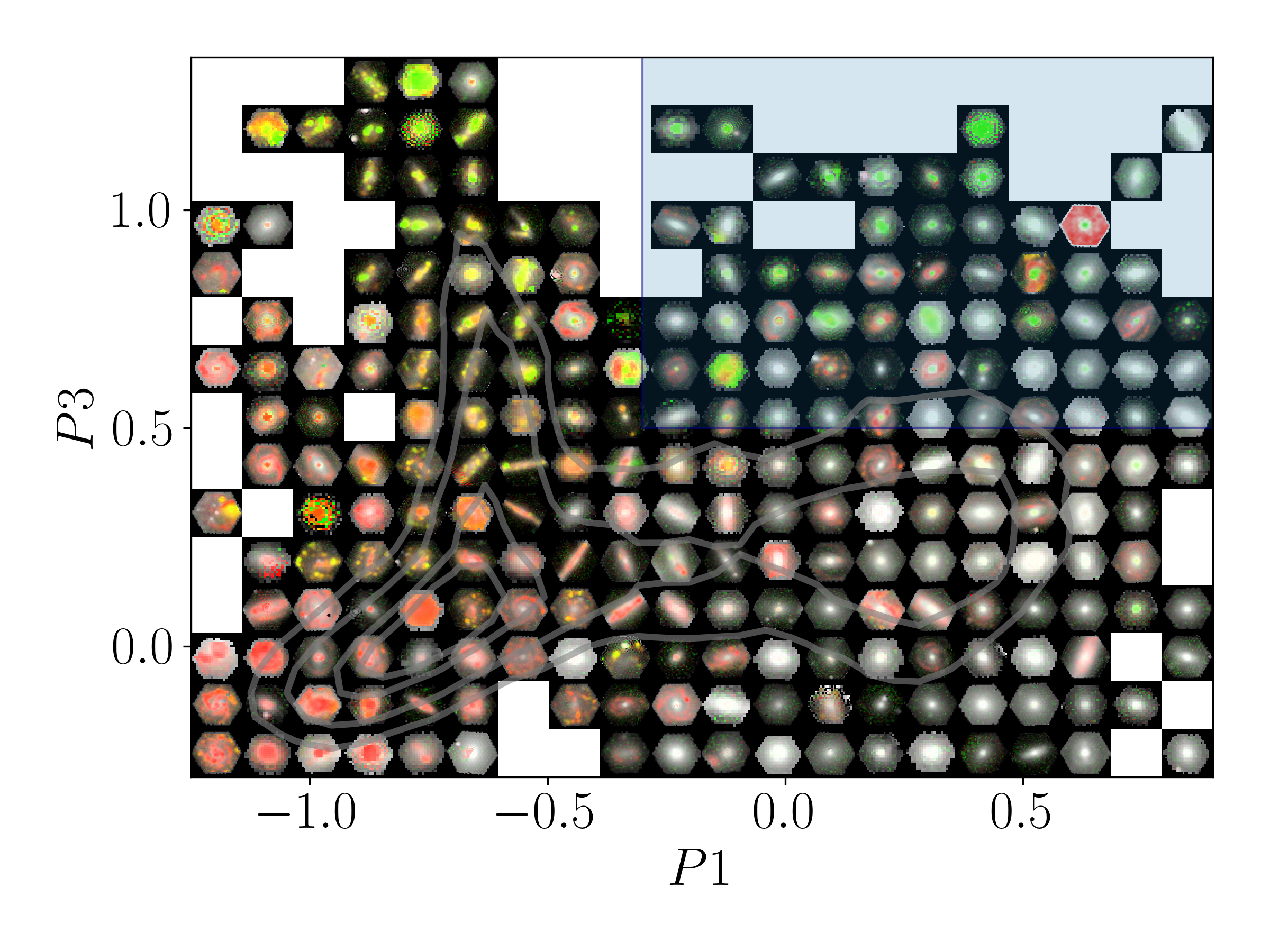}
\end{center}
\caption{\label{fig:p1p3} Line ratio diagnostic diagram of
  \citet{ji20a}, showing $P1$ and $P3$ for MaNGA galaxy central
  emission-line ratios. The left panel shows the measured values
  and errors, with the distribution overlaid.  The right panel
  shows one randomly selected galaxy at each location in the
  plane. The $r$-band image is shown as a gray scale, with the
  \ha\ map contributing to the red channel and the \oiiilam\ map
  contributing to the green channel. In both panels, the blue box
  shows our Seyfert galaxy classification.}
\end{figure}

The left panel of Figure \ref{fig:p1p3} shows the distribution of
MaNGA galaxy central emission-line ratios in the space of $P1$ and
$P3$; only the 7,672 galaxies with all relevant emission lines
detected at $2\sigma$ are shown. The galaxies with emission-line
ratios dominated by star formation form the sequence on the left. The
LINER-like galaxies are on the bottom right. The Seyfert galaxies are
on the upper right, and our selection is shown as the blue box. The
right panel of Figure \ref{fig:p1p3} shows the image of one galaxy at
each corresponding $P1$--$P3$ value. The $r$-band image is shown as
gray scale, and the \ha\ map is overlaid in red and the \oiiilam\ map
is overlaid in green. The metallicity sequence on the right
(increasing from top to bottom) is visible as the images go from
\oiii\ to \ha\ dominated.  The LINER \oiii\ emission tends to be
subdominant to the galaxy, but for Seyferts it is more visible (see
the right panel of Figure 15 of \citealt{ji20a} showing the
\ha\ equivalent widths as a function of $P3$).

There are 306 detected Seyfert galaxies in the full MaNGA sample by
these definitions, 265 in the statistical sample (Primary,
Color-Enhanced, and subsampled Secondary samples), and 210 of those
have $\sigma_e> 60$ km~s$^{-1}$. For each detected Seyfert we assume the
central \oiiilam\ luminosity ($\loiii$) and \hb\ luminosity ($\lhb$)
reflect mostly the light from a central AGN source (we discuss dust
corrections in Section \ref{sec:bolometric}).

\begin{figure}
\begin{center}
  \includegraphics[width=0.98\textwidth]{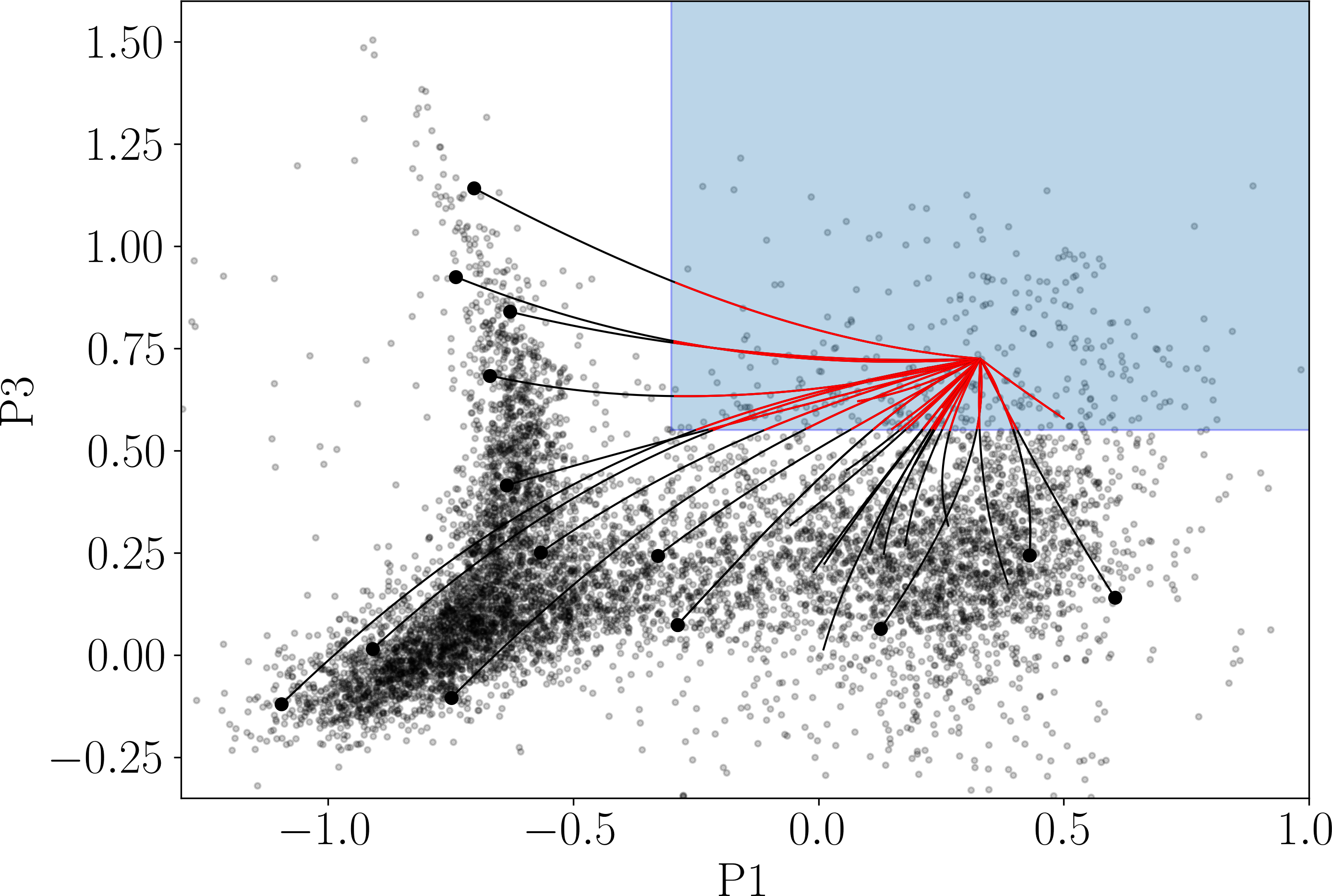}
\end{center}
\caption{\label{fig:p1p3_selection} Same line ratio diagnostic diagram
  as in Figure \ref{fig:p1p3}, showing the same galaxies, illustrating
  our determination of \loiii\ detection thresholds for non-Seyfert
  galaxies. The blue region shows the line ratios for which we define
  Seyfert galaxies. The small gray points show each galaxy with all
  five lines detected at at least 2$\sigma$ significance (error bars
  are omitted for clarity). The large black points are a random set of
  non-Seyfert galaxies chosen to span a range of metallicities and
  ionization spectral shapes (i.e. $P3$ and $P1$ values). The lines
  emanating from each large black point show how the line ratios
  change with an addition of an AGN component of increasing
  luminosity. Note that some lines exist without a large black point;
  those lines are examples for which the host galaxy's lines are not
  all detected at 2$\sigma$, and the lines start at AGN
  luminosities that would be necessary to make the lines
  detectable. The lines are colored red for AGN luminosities high
  enough to be identified as AGN by our methods and black otherwise.
}
\end{figure}

\subsection{Determining Identification Thresholds}
\label{sec:agn-upper-limits}

The galaxies in the sample (the majority) that are not identified as
Seyfert are best understood as not having high enough Seyfert
emission to be detectable. Many of them could, and most likely do,
host lower-luminosity Seyfert emission surrounding a central
supermassive black hole that does not perturb the central line ratios
enough to be detectable. Following the methodology of
\citet{trump15a}, we determine detection thresholds for the
luminosity of the Seyfert galaxy that could be in each galaxy yet
remain undetectable according to our threshold estimates.

For each galaxy, we take its central line fluxes and their errors and
we add to them a template set of line fluxes with nominal Seyfert-like line
ratios:
\begin{eqnarray}
\label{eq:nominal}
\log_{10} \oiiihb &=& 0.779 \cr
\log_{10} \niiha &=& -0.017\cr
\log_{10} \siiha &=& -0.233
\end{eqnarray}
which are equivalent to $P_1 = 0.330$, $P_2=-0.171$, and $P_3=0.725$.

Scaling these according to \loiii, we then find the AGN luminosity
\loiii\ that would perturb the line ratios enough to both exceed the
$2\sigma$ detection limit in each line and cross into our Seyfert
selection.  Figure \ref{fig:p1p3_selection} illustrates the technique
for a subset of the galaxies. The lines show the path in $P1$--$P3$
space as the AGN luminosity is increased.  Note that some lines
originate from a large black point on the plot (those galaxies that
already have detected lines at 2$\sigma$), and some lines do not
(those galaxies for which one or more lines are not detected already).

This process yields an AGN luminosity detection threshold in
\llimoiii\ and \llimhb\ (and similar quantities for the dust-corrected
luminosities) for each galaxy in our sample that does not have an
actual detection.  We use these thresholds below to help
constrain the Eddington ratio distributions.

\subsection{Eddington Ratio Determination}
\label{sec:agn-derived}

\subsubsection{Eddington Ratio Definition}
\label{sec:eddington_ratio}

The Eddington ratio $\lambda$ is defined as the ratio of the bolometric
luminosity to the Eddington luminosity, where the latter is the
luminosity that would prevent gravitational accretion of hydrogen gas
in a spherically symmetric flow, which is $L_{\rm Edd} = 1.38 \times
10^{38} (M_{\rm BH} / M_\odot)$ erg~s$^{-1}$.  The Eddington ratio is
therefore closely related to the observed luminosity divided by the
black hole mass. As such, it provides a reasonable characterization of
the level of ``activity'' associated with the black hole accretion.
However, as defined it requires converting the observed luminosity in
the AGN signature, in our case \loiii\ or \lhb, to bolometric
luminosity, and it requires an inference of the black hole mass. In
this section we describe these two uncertain inferences.

\subsubsection{Bolometric Luminosities}
\label{sec:bolometric}

Determining $L_{\rm bol}$ is fraught with systematic issues. We 
use several different determinations to quantify the nature of the
errors these issues might cause in our analysis. We consider
conversions that are observationally based on the relationship between
narrow lines and bolometric luminosity for Type 1 AGN, whose
panchromatic emission can be measured along with the narrow lines
(\citealt{heckman04a, kauffmann09a}), as well as the theoretical model
of \citet{netzer19a}.

An initial consideration is dust extinction. For a narrow-line AGN,
there is dust associated with the gas in the narrow-line region
itself, and there is dust in the host galaxy's interstellar medium
that the light may have to pass through. With the data available for
our sample, we cannot disentangle these effects. When we need to
correct for dust extinction, we use a Balmer-decrement-based
approach. We measure the $\ha/\hb$ flux ratio, which assuming a
universal attenuation curve then determines the absolute level of
attenuation. For this purpose, we assume the attenuation curve of
\citet{whitford58a} as parameterized by \citet{miller72a}.  For those
galaxies without measured \ha\ or \hb\ lines, even though they are not
detected AGN (since they do not have all of the necessary lines
detected), we still need to calculate a dust correction to calculate
the dust-corrected luminosity detection threshold. For this purpose,
we use the median dust correction for the detected AGN ($A_{{\rm
    H}\alpha} = 0.949$ mag).

\citet{kauffmann09a} apply a Balmer-decrement-based dust correction
to \loiii, producing a dust-corrected luminosity \loiiicorr. They then
apply a factor of 600 to obtain a bolometric luminosity:
\begin{equation}
\label{eq:kauffmann}
\log_{10} L_{\rm bol} = 2.778 + \log_{10} \loiiicorr
\end{equation}
They chose this conversion because it was roughly between the
corrections cited by \citet{ho08a} for lower-luminosity AGN and those
cited by \citet{lamassa09a} (referred to as an in preparation paper in
\citealt{kauffmann09a}) for higher luminosity AGN. \citet{ho08a}
quotes an \ha-based bolometric correction, which they state is based
on a Ho et al. (in preparation) paper, which may be related to
\cite{ho09a}, though the latter does not quote any bolometric
corrections.  Regardless, \citet{kauffmann09a} convert the \ha-based
correction to \oiii, and find that the bolometric correction factor is
300--600. \citet{lamassa09a} calculate the bolometric correction
factor for higher-luminosity AGN as 500-900 based on comparisons to
mid-infrared luminosities.  \citet{kauffmann09a} motivate the choice
of their factor of 600 as being approximately between these two
ranges.

An alternative conversion from \citet{heckman04a} uses uncorrected
\oiii\ luminosities:
\begin{equation}
\label{eq:heckman}
\log_{10} L_{\rm bol} = 3.544 + \log_{10} \loiii
\end{equation}
They derive this conversion from the ratio of optical continuum to
\oiiilam\ luminosity for broad-line quasars (\citealt{kauffmann03b,
  zakamska03a}) and the conversion from optical continuum to
bolometric from \citet{marconi04a}. Because in their first step they
do not correct either the line or continuum luminosity for dust, the conversion applies to
uncorrected luminosities. The difference between Equation
\ref{eq:heckman} and Equation \ref{eq:kauffmann} is about 0.2 dex (or
about 0.5 mag) larger than the difference we would infer from the
median dust correction quoted above.

Whether to base a bolometric luminosity on dust-corrected luminosities
or not is unclear.  The ratio between the optical \oiii\ and the
mid-infrared \oiv\ lines suggests that in Type 1 AGN the narrow lines
experience less dust attenuation than do Type 2 sources
(\citealt{diamondstanic09a}), by a factor similar to the difference
between the conversions in Equation \ref{eq:kauffmann} and
\ref{eq:heckman} (i.e. 1--2 mag). So the two conversions appear
roughly consistent, and applying the dust correction should lead to
higher accuracy. Yet \citet{lamassa10a} show that correlations between
\oiii\ and presumably less dust-affected indicators like the
mid-infrared luminosity, \oiv, and hard X-rays are better with
\loiii\ than with \loiiicorr. 

The bolometric correction may also depend on black hole mass,
especially for \loiii, due to the changing spectrum of the accretion
disk predicted by models like those of
\citet{shakura73a}. \citet{cann19a} explore this dependence and find
that at high masses ($M_{\rm BH} \sim 10^8$) most of the oxygen is
singly  or doubly ionized, that as black hole mass decreases more of
the oxygen is more highly ionized, and that at a given accretion rate in
their models \loiii\ is correlated with black hole mass.

Whether to account for a luminosity dependence in the bolometric
correction is also unclear. \citet{ho08a}, \citet{kauffmann09a}, and
others claim that the bolometric corrections rise with luminosity.
Theoretically, \citet{netzer19a} and \citet{cann19a} predict
bolometric corrections that rise with luminosity.

\citet{netzer19a} makes a specific prediction for \hb; he does not
perform the calculations necessary to make a reliable prediction for
\oiii. His prediction is
\begin{equation}
\label{eq:netzer}
\log_{10} L_{\rm bol} = 45.661 + 1.18 \left(\log_{10} \lhbcorr -
42\right),
\end{equation}
(where we write \lhbcorr\ to indicate that this luminosity needs to be
dust corrected to use this theoretical model).  Additionally,
\citet{netzer19a} find that black hole mass and spin variations
produce anywhere from 0.2 to 1 dex in peak-to-peak variations of \lhb,
depending on the bolometric luminosity. These models therefore predict
considerable scatter in the bolometric corrections.

%

In short, bolometric corrections from narrow-line luminosities are
highly uncertain, despite extensive efforts to determine them. We
derive three different bolometric luminosities meant to span the space
of possibilities: (a) \loiii\ uncorrected for internal dust
extinction and apply Equation \ref{eq:heckman} to \loiii; (b) 
\loiiicorr\ corrected for internal dust extinction using the Balmer
decrement and apply Equation \ref{eq:kauffmann}; and (c) 
\lhbcorr\ corrected for internal dust extinction using the Balmer
decrement and apply Equation \ref{eq:netzer}. Our figures below show
the results for the \lhbcorr-based bolometric luminosities, but the
differences among these approaches turn out to not affect our
qualitative results.

\subsubsection{Black hole masses}
\label{sec:bhmasses}

Estimating black hole masses is similarly fraught, and for our sample
these masses cannot be obtained directly. Instead we use the
relationship between black hole mass and stellar velocity dispersion
($M_{\rm BH}$--$\sigma$) calibrated using high angular resolution
stellar or gas dynamics for nearby galaxies (see \citealt{kormendy13a}
and references therein). We use three versions of this
relationship to test the sensitivity of our results to this choice.

For the stellar velocity dispersion, we use $\sigma_e$ within a
half-light radius as estimated by the MaNGA DAP. As noted above, we
will use $\sigma_e>60$ km~s$^{-1}$. This quantity is not precisely the
quantity used by most $M_{\rm BH}$--$\sigma$ calibrations, which
usually calculate $\sigma^2$ from the light-weighted average of $V^2
+ \sigma^2$ within the half-light radius, where $V$ is the mean
stellar velocity in the galaxy's frame. At the level of accuracy we
are working, we expect this difference to be a small effect, and in
any case the physical resolution available for MaNGA makes a
central estimate of this quantity difficult.

The form of $M_{\rm BH}$--$\sigma$ is usually taken to be a power law:
\begin{equation}
\label{eq:msigma}
\log_{10} M_{\rm BH} = \alpha + \beta \log_{10}
\left(\frac{\sigma}{200 {\rm ~km~s}^{-1}}\right)
\end{equation}
with $\beta \sim$ 4--6 depending on the analysis. \citet{kormendy13a}
find $\beta = 4.4 \pm 0.3$ and $\alpha = 8.50\pm 0.05$ (specifically,
their Equation 5). \citet{graham13a} find $\beta = 5.2 \pm 0.3$ and
$\alpha = 8.14 \pm 0.04$ (we use their regression given at the end of
their Section 3.1 of black hole mass on $\sigma$ for barred and
unbarred galaxies together, which of their methods is the most
appropriate one for inferring the $M_{\rm BH}$ from
$\sigma$). Finally, fits to this relationship determined only for
local AGN find a lower zero-point and shallower slope, for example
$\alpha = 7.86 \pm 0.04$ and $\beta = 3.65 \pm 0.13$
(\citealt{greene06a}).\footnote{Note that these numbers
come from the preprint {\tt v1} of \citet{greene06a}, whereas in the
final published version of \citet{greene06a}, the corresponding fit is
$\alpha = 7.85 \pm 0.04$ and $\beta = 3.69 \pm 0.13$; we used the
preprint numbers in error, but we caught this discrepancy only at a
late stage of our analysis, and given that it is well within the
uncertainties, we have not recalculated our results.}

The samples of non-AGN galaxies that lead to different values for
$\alpha$ and $\beta$ samples tend to have overlapping samples, but the
values are affected by precisely which sample is used
(\citealt{park12a}). The flattening and shift of the AGN-based sample
may be affected by the inability to measure black hole masses
dynamically for the highest-mass black holes when there is an AGN
(\citealt{woo13a}). 

In short, the black hole masses are rather uncertain. We use the
three relations quoted above (\citealt{kormendy13a, graham13a,
  greene06a}), and evaluate whether the exact adopted black hole
masses matter, which qualitatively they do not.

\section{Luminosities and Eddington Ratios}
\label{sec:luminosities}

\begin{figure}[t!]
\begin{center}
  \includegraphics[width=0.98\textwidth]{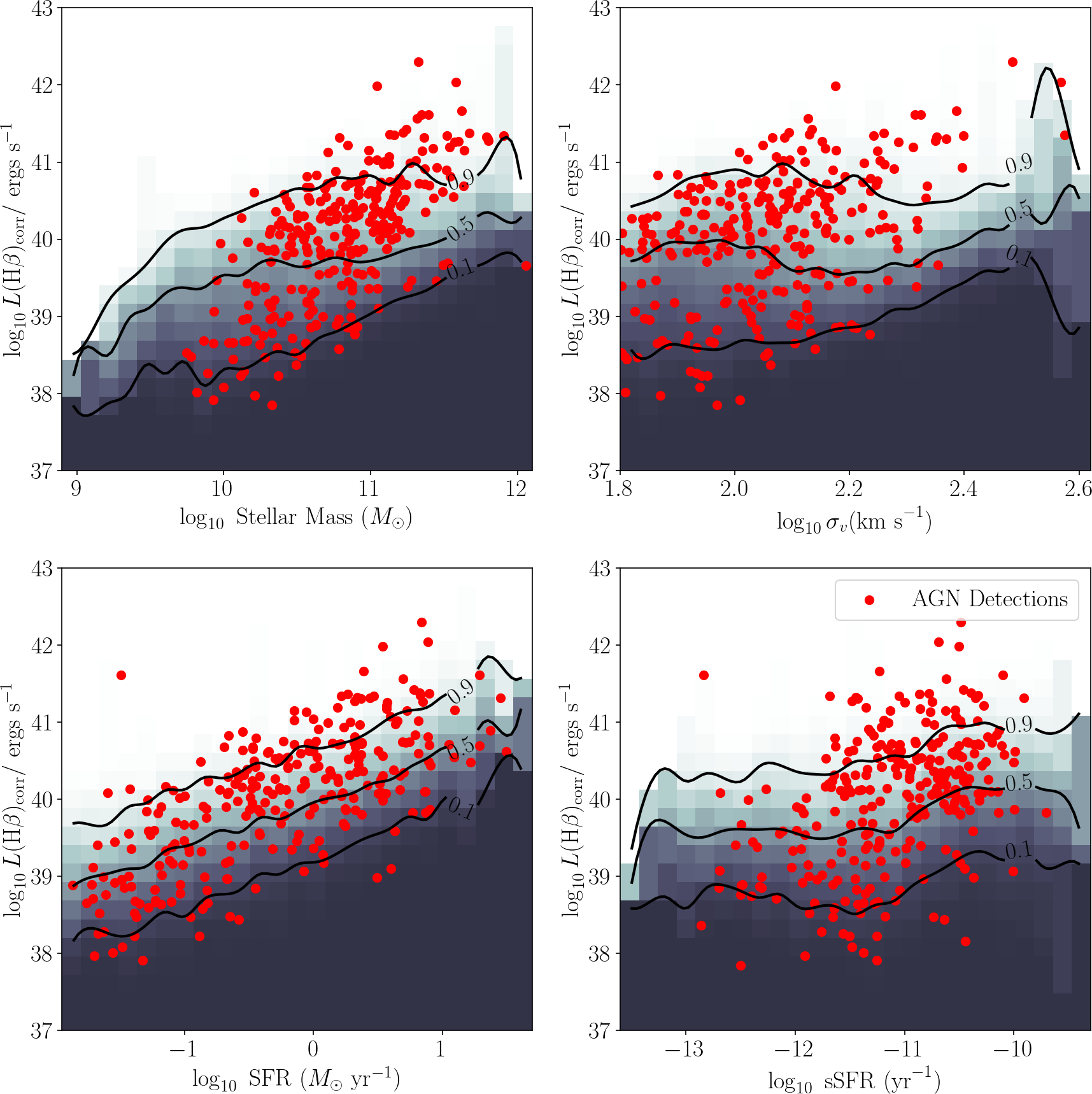}
\end{center}
\caption{\label{fig:hb_lums_and_thresholds} AGN-related dust-corrected
  H$\beta$ luminosities and detection thresholds for MaNGA galaxies,
  as a function of various host galaxy properties: stellar mass (top
  left), stellar velocity dispersion $\sigma_e$ (top right), SFR
  (bottom left), and sSFR (bottom right). The red points
    show detections. The vast majority ($\sim 97\%$) of galaxies do
    not have detections and yield only luminosity thresholds.  The
    gray scale shows the cumulative conditional distribution of these
    thresholds. That is, for each value of each host galaxy property,
    it shows the fraction of galaxies for which any given luminosity
    would be detectable. For example, at low SFR, an AGN with
    $\lhbcorr \sim 10^{40}$ erg~s$^{-1}$ would be detectable for 90\%
    of galaxies, but at high SFR the same AGN would be detectable only
    for 10\% of galaxies. }
  
\end{figure}

Figure \ref{fig:hb_lums_and_thresholds} shows the dust-corrected
\hb\ luminosities for the AGN, and the cumulative conditional
distribution of luminosity thresholds for galaxies without detected
AGN. We show the dependence on stellar mass, velocity dispersion
$\sigma_e$, SFR, and sSFR.  Although we plot the values and detection
thresholds for the dust-corrected \hb\ luminosity \lhbcorr\ here, the
same analysis produces results for \lhb, \loiii, and \loiiicorr.

This figure illustrates the difficulty in interpreting the narrow-line
AGN population statistics. Only a handful of narrow-line AGN are so
luminous that they would be detectable in all galaxies. For a typical
identified AGN luminosity of $\lhbcorr\sim 10^{40}$ erg~s$^{-1}$, it
would be detectable in about 65\% of the galaxies.  As Figure
\ref{fig:hb_lums_and_thresholds} demonstrates, the detectability is a
strong function of host galaxy properties. Therefore, the shape of the
luminosity distribution and its dependence on host galaxy properties
must be dramatically affected by these selection effects.

\citet{trump15a} and \citet{jones16a} have both explored these
detection thresholds in detail before, in the context of the SDSS
Legacy Main Sample of galaxies (\citealt{strauss02a}). Comparing to
Figure 8 of \citet{trump15a}, we find a similar range of detection
thresholds when expressed in the units they use, with $\log_{10}
\loiiicorr / M_\ast \sim 29$--$31$, expressing the ratio in units of
ergs~s$^{-1}$~$M_\odot^{-1}$.

\citet{heckman04a} also analyzed the completeness effects in the same
SDSS Legacy Main Sample, claiming completeness above $\loiii \approx
10^6$ $L_\odot \sim 4\times 10^{39}$ ergs~s$^{-1}$ (not dust
corrected) for all potential host galaxies.  Figure 8 in
\citet{trump15a} shows that much of the host galaxy parameter space
actually has a higher threshold for detection. For example, accounting
for a typical dust correction of 0.6 dex, a star forming galaxy with
$\log_{10} M_\ast \sim 10$ has a threshold of $\loiii \sim 10^{40}$
ergs~s$^{-1}$ and a galaxy at $\log_{10} M_\ast \sim 11$ has a
threshold of $\loiii \sim 10^{41}$ ergs~s$^{-1}$, much higher than
\citet{heckman04a} claim. The numbers from our analysis are not
directly comparable because we use a different dataset and selection
method, but our conclusions are similar to those of \citet{trump15a},
with only 20\% of potential host galaxies in the MaNGA sample having
thresholds as sensitive as \citet{heckman04a} claim for all galaxies.
Our conclusion is that the claimed completeness limits of
\citet{heckman04a} cannot be used to accurately infer AGN host galaxy
population properties in either the SDSS Legacy Survey Main Sample or
the MaNGA sample.

\begin{figure}[t!]
\begin{center}
  \includegraphics[width=0.98\textwidth]{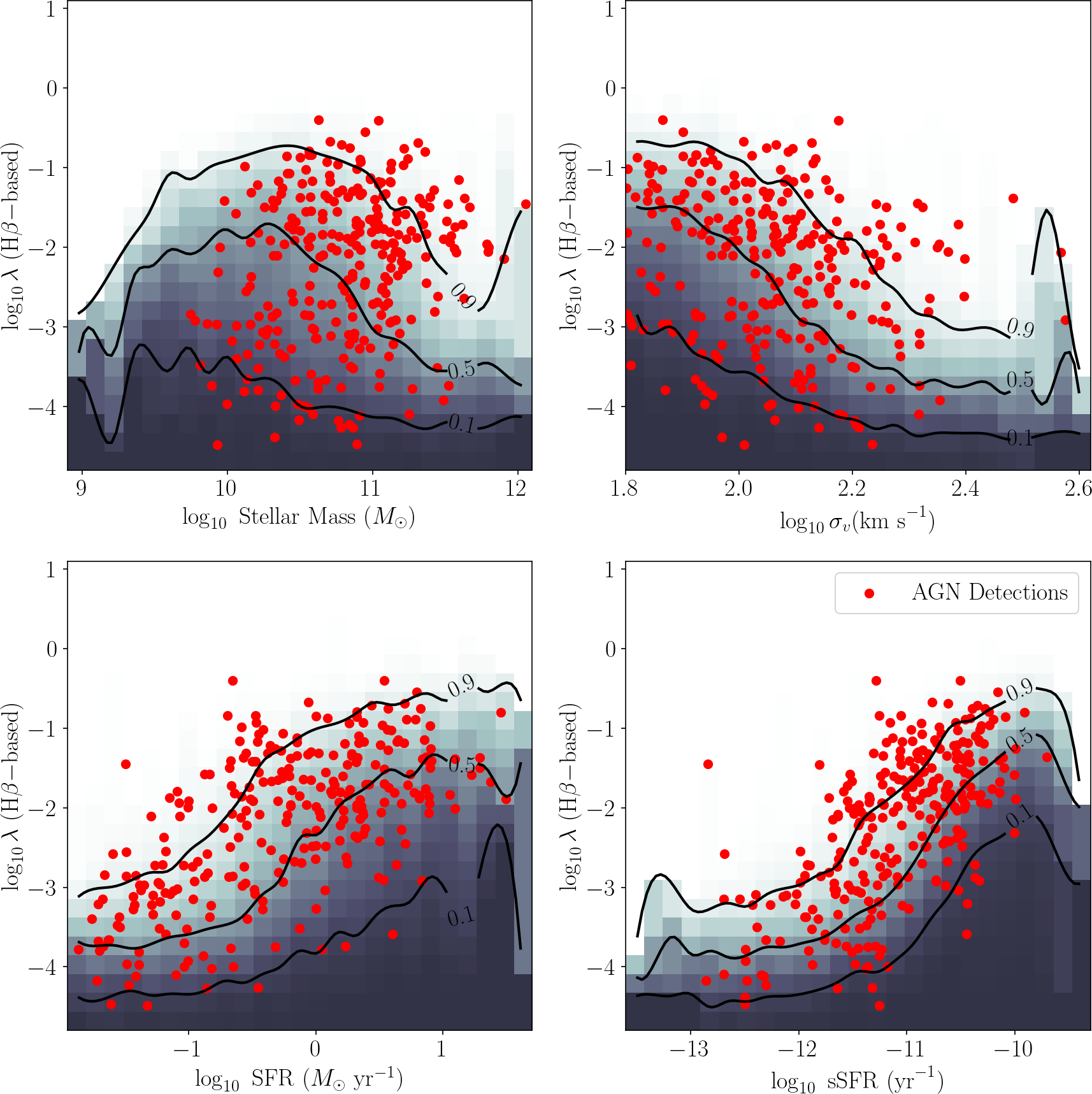}
\end{center}
\caption{\label{fig:ers_and_thresholds_netzer_kormendy} Similar to
  Figure \ref{fig:hb_lums_and_thresholds}, for Eddington ratios and Eddington
  ratio detection thresholds.  We use the \citet{kormendy13a} $M_{\rm
    BH}$--$\sigma$ relation and the \citet{netzer19a} bolometric
  corrections to calculate the Eddington ratios; the qualitative
  trends do not depend on these choices.}
\end{figure}

Figure \ref{fig:ers_and_thresholds_netzer_kormendy} shows the Eddington
ratios and Eddington ratio detection thresholds derived from \lhb,
using the \citet{kormendy13a} $M_{\rm BH}$--$\sigma$ relation and the
\citet{netzer19a} bolometric corrections. These values look as
expected given Figure \ref{fig:hb_lums_and_thresholds}, again illustrating
that the selection thresholds must play an important role in shaping the
distribution of identified AGN luminosities and of their host galaxy
properties.

We point out two features in particular. First, the detection
thresholds in Eddington ratio become quite high at low velocity
dispersion (and to a lesser extent, at low mass). This feature is
caused by the strong assumed dependence of black hole mass on velocity
dispersion; although the \lhb\ thresholds are not a strong function of
$\sigma_e$ (Figure \ref{fig:hb_lums_and_thresholds}), these thresholds
correspond to much higher Eddington ratios at low $\sigma_e$.

Second, at high sSFR, the thresholds tend to exceed the luminosities of
most detected AGN. This effect is mostly due to the greater degree of
contamination by star formation related narrow-line emission at high
sSFR, and partly due to an anti-correlation of sSFR and $\sigma_e$,
which therefore leads to higher Eddington ratios at higher sSFR.

\begin{figure}[t!]
\begin{center}
  \includegraphics[width=0.98\textwidth]{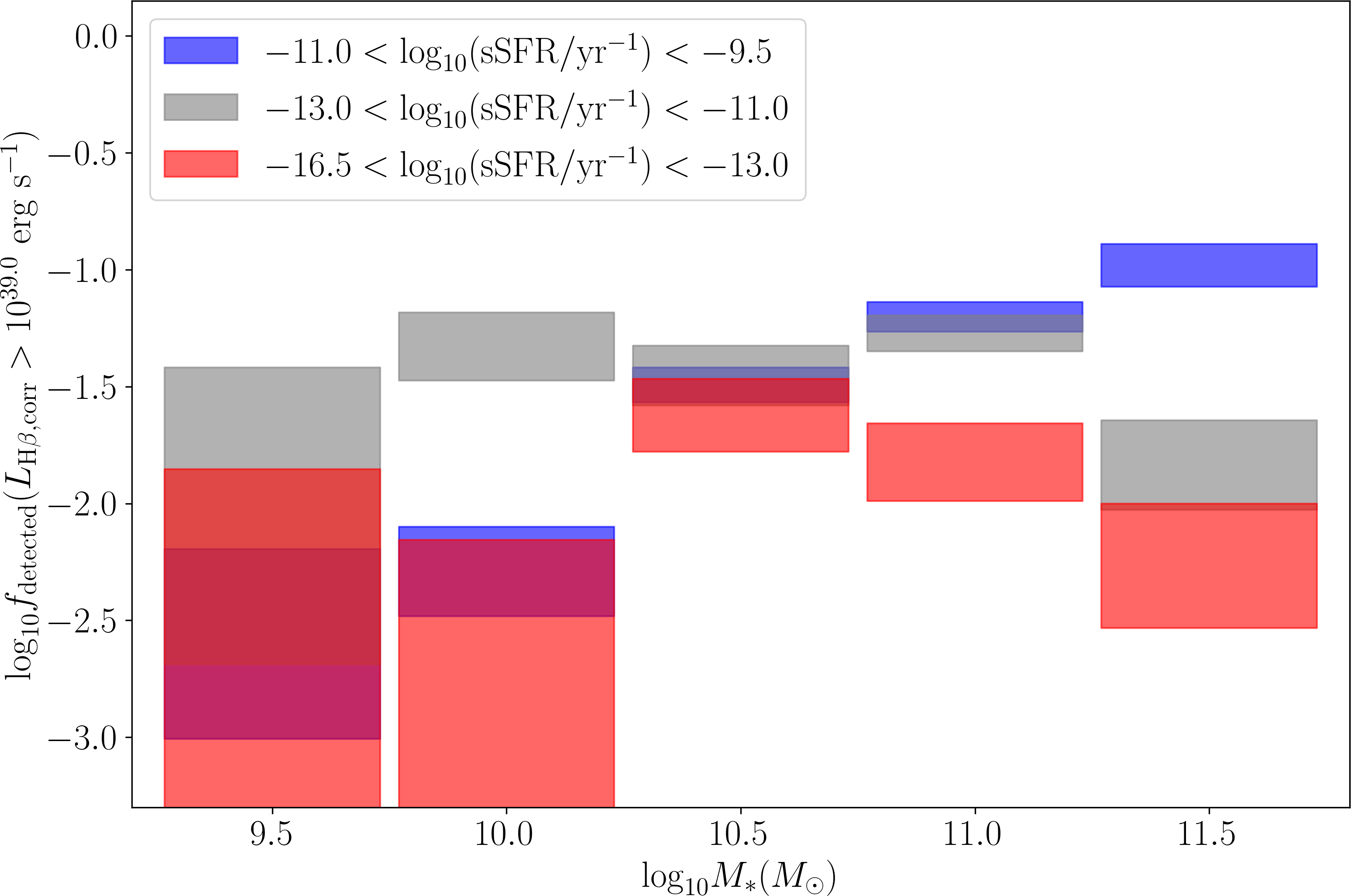}
\end{center}
\caption{\label{fig:lumhb_mssfr_raw} AGN raw detection rate $f_{\rm
    detected}(\lhbcorr > 10^{39}$ erg~s$^{-1})$ as a function of
  stellar mass and sSFR. These detection rates have no correction for
  the rather severe selection effects and therefore should not be
  taken at face value. The bands show 68\% confidence intervals (using
  the Clopper-Pearson method for binomial statistics;
  \citealt{clopper34a}).}
\end{figure}

\begin{figure}[t!]
\begin{center}
  \includegraphics[width=0.98\textwidth]{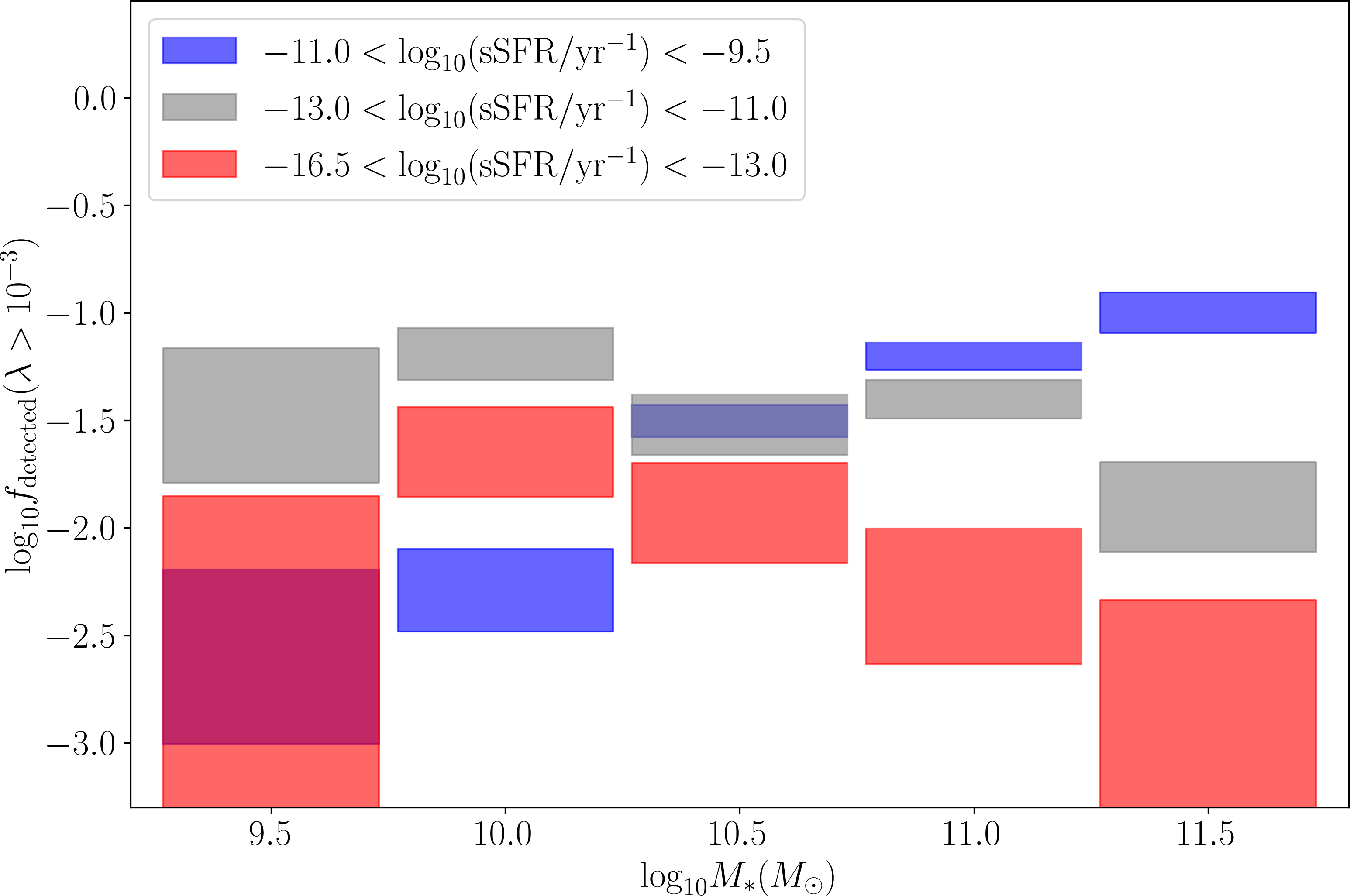}
\end{center}
\caption{\label{fig:edr_mssfr_raw} Similar to Figure
  \ref{fig:lumhb_mssfr_raw}, for Eddington ratio selected AGN with
  $f_{\rm detected}(\lambda > 10^{-3})$. These detection rates have no correction for the rather
  severe selection effects and therefore should not be taken at face
  value.}
\end{figure}

These sorts of effects complicate the interpretation of the
distribution of host galaxy properties of AGN.  However, it is useful
to consider the distribution AGN detections as a function of stellar
mass and sSFR, before any correction for these effects. Figures
\ref{fig:lumhb_mssfr_raw} and \ref{fig:edr_mssfr_raw} show these
fractions for luminosity-selected and Eddington ratio-selected AGN,
respectively. There is a strong stellar mass dependence for
star-forming galaxies. Both intermediate and high sSFR galaxies show
an enhancement in AGN relative to quiescent galaxies, with about the
same fraction at $M_\ast \sim 10^{10} M_\odot$, and a more dramatic
correlation with sSFR at higher masses. However, these results do not
reflect the total AGN fraction because of the severe selection effects
on these detection rates.

\section{Narrow-line Luminosity Distributions}
\label{sec:ldist}

\subsection{Methodology}
\label{sec:methodology_lum}

In this section we estimate the narrow-line luminosity distributions
for \loiii, \loiiicorr, \lhb, and \lhbcorr, as a function of mass and
sSFR. Specifically, we estimate the probability distribution
$\Phi(L)$, as well as the conditional probability distributions
$\Phi(L|M_\ast, {\rm sSFR})$; these quantities are expressed
per-unit-luminosity and integrate to unity. Our primary analysis uses
a parametric approach to this estimate, which we validate using a
nonparametric approach.

In our primary analysis, the parametric function we use for the
probability distribution is the Schechter function:
\begin{equation}
\label{eq:schechter}
\Phi(L) = \left\{
\begin{array}{cc}
  \phi_0 \left(\frac{L}{L_\ast}\right)^{- \alpha} \exp\left(- L /
  L_\ast\right) & {\rm if~}L > L_{\rm min} \cr
  0 & {\rm if~}L \le L_{\rm min}
  \end{array} \right. 
\end{equation}
where $\Phi(L)$ is per unit luminosity, and there is a minimum
luminosity $L_{\rm min}$ below which $\Phi(L)=0$. In Equation
\ref{eq:schechter}, $\phi_0$ is in units of inverse luminosity and is
defined such that the integral of $\Phi(L)$ is unity, so the free
parameters of the model are $\alpha$, $L_\ast$, and $L_{\rm min}$.

As we show below, with our dataset the individual parameters of
this Schechter function are highly covariant and not particularly
well-constrained. However, based on a given set of parameters we can
define $F_{\rm AGN, Y}$ as the fraction of galaxies with AGN above a
particular luminosity $L_Y$, where $Y = \log_{10} (L_Y / ({\rm
  erg}~{\rm s}^{-1}))$. It turns out that for a choice of $Y \sim
39$--$40$, near the typical luminosity thresholds seen in Figure
\ref{fig:hb_lums_and_thresholds}, $F_{\rm AGN, Y}$ is well constrained.

For any sample or subsample of galaxies with luminosities $L$ and
threshold luminosities $L_{\rm lim}$ we can define the log-likelihood
of the data as 
\begin{equation}
  \ln P(\{L_i\}, \{L_{{\rm lim}, j}\} | \alpha, L_\ast, L_{\rm min}) = \left(\sum_i
  \ln \Phi(L_i) \right) +
  \left(\sum_j \ln \int_{L_{\rm min}}^{L_{{\rm lim}, j}} {\rm d}L \Phi(L)
    \right),
\end{equation}

To constrain $\alpha$ and $L_\ast$ for any sample or subsample, we
take a Bayesian approach and perform Markov Chain Monte Carlo using
{\tt emcee} (\citealt{foremanmackey13a}). We assume a uniform prior on
$\alpha$, $\log_{10} L_\ast$, and $\log_{10} L_{\rm min}$ over a
finite range in each value. We limit the slope to $-1.9 < \alpha <
2.4$, and for $\log_{10} L_\ast$ and $\log_{10} L_{\rm min}$ we limit
to ranges that depend on the exact sample (designed to be less than
the minimum observed luminosity or luminosity threshold).

We validate our choice of the Schechter function parameterization
using the nonparametric statistic of \citet{kaplan58a}. In this
approach, we estimate the cumulative distribution expressing the
fraction of galaxies with an AGN above luminosity $L_c$,
\begin{equation}
    F_{\rm AGN}(>L_c) = \int_{L_c}^\infty {\rm d}L \Phi(L),
\end{equation}
using the estimator
\begin{equation}
  \hat F_{\rm AGN}(>L_c) = 1 - \prod_{i: L_i > L_c} \left[1 - \frac{1}{N(<
      L_i)}\right].
\end{equation}
In this equation, $i$ indexes all the detected AGN and the product
includes all those AGN with luminosities greater than $L_c$.
$N(<L_i)$ indicates the number of galaxies with AGN detection
thresholds smaller than $L_i$ and no detected AGN, plus all detected
AGN with luminosities smaller than $L_i$. Thus $1-1/N(<L_i)$ can be
interpreted as the fractional decrement in the cumulative probability
distribution $F_{\rm AGN}(L)$ caused by the detection of an AGN with
$L=L_i$ (because the galaxy at $L_i$ constitutes a fraction
$1/N(<L_i)$ of the galaxies that could have had an AGN detected at
that luminosity). The variance in this estimate can be estimated as
\begin{equation}
  {\rm Var}\left(\hat F_{\rm AGN}\right) = \left(1 - \hat F_{\rm
    AGN}\right)^2 \sum_{i:L_i > L_c} \frac{1}{N(<L_i) \left(N(<L_i)
    - 1\right)},
\end{equation}
using the formula of \citet{greenwood26a}. 

\subsection{Full sample analysis of the $\hb$ luminosity distribution}
\label{sec:lum_uniform}

To illustrate the results, we first consider the full sample,
undifferentiated by stellar mass or sSFR. We fit the distribution
of the dust-corrected \hb\ luminosity, \lhbcorr.

\begin{figure}[t!]
\begin{center}
  \includegraphics[width=0.48\textwidth]{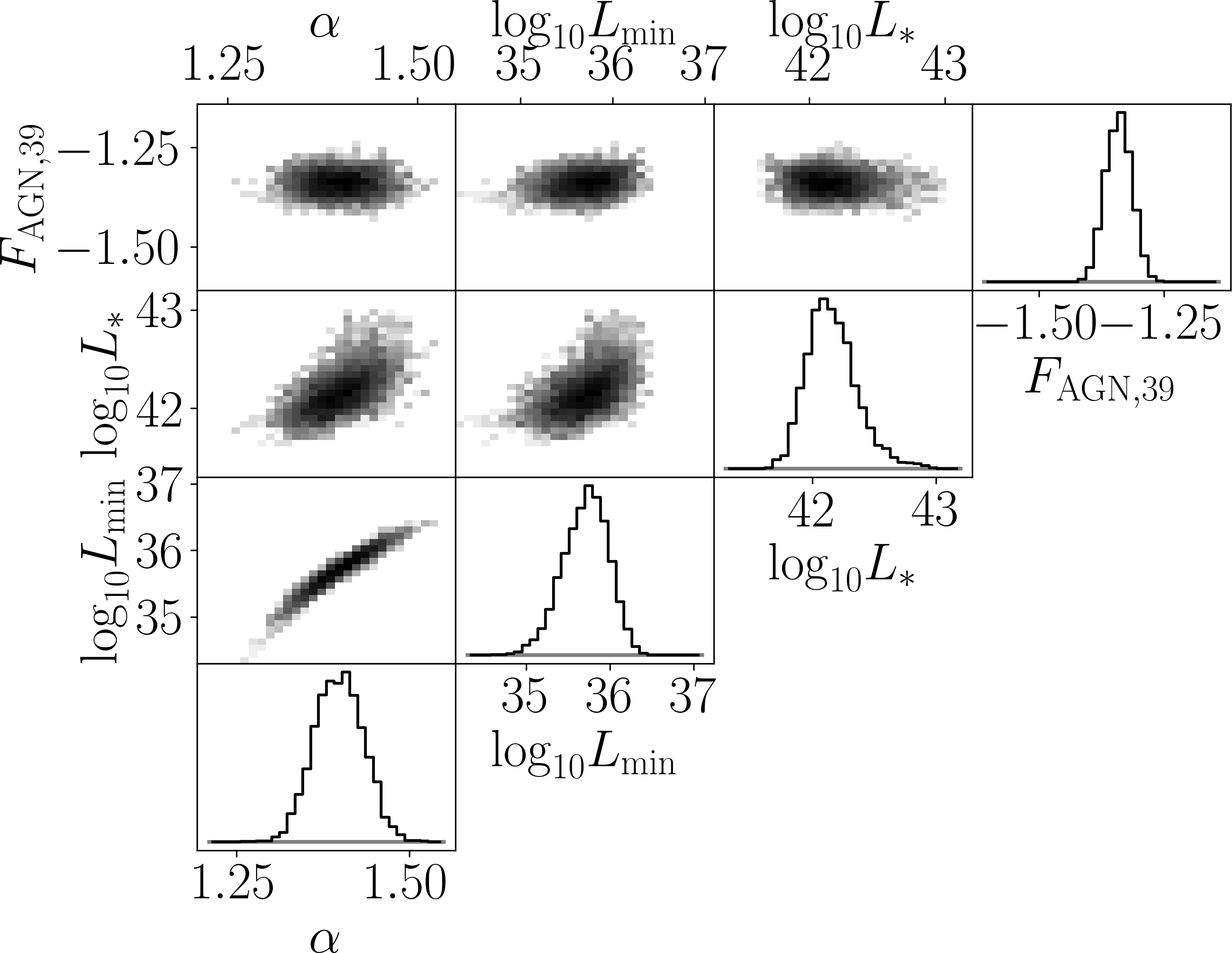}
  \quad
  \includegraphics[width=0.48\textwidth]{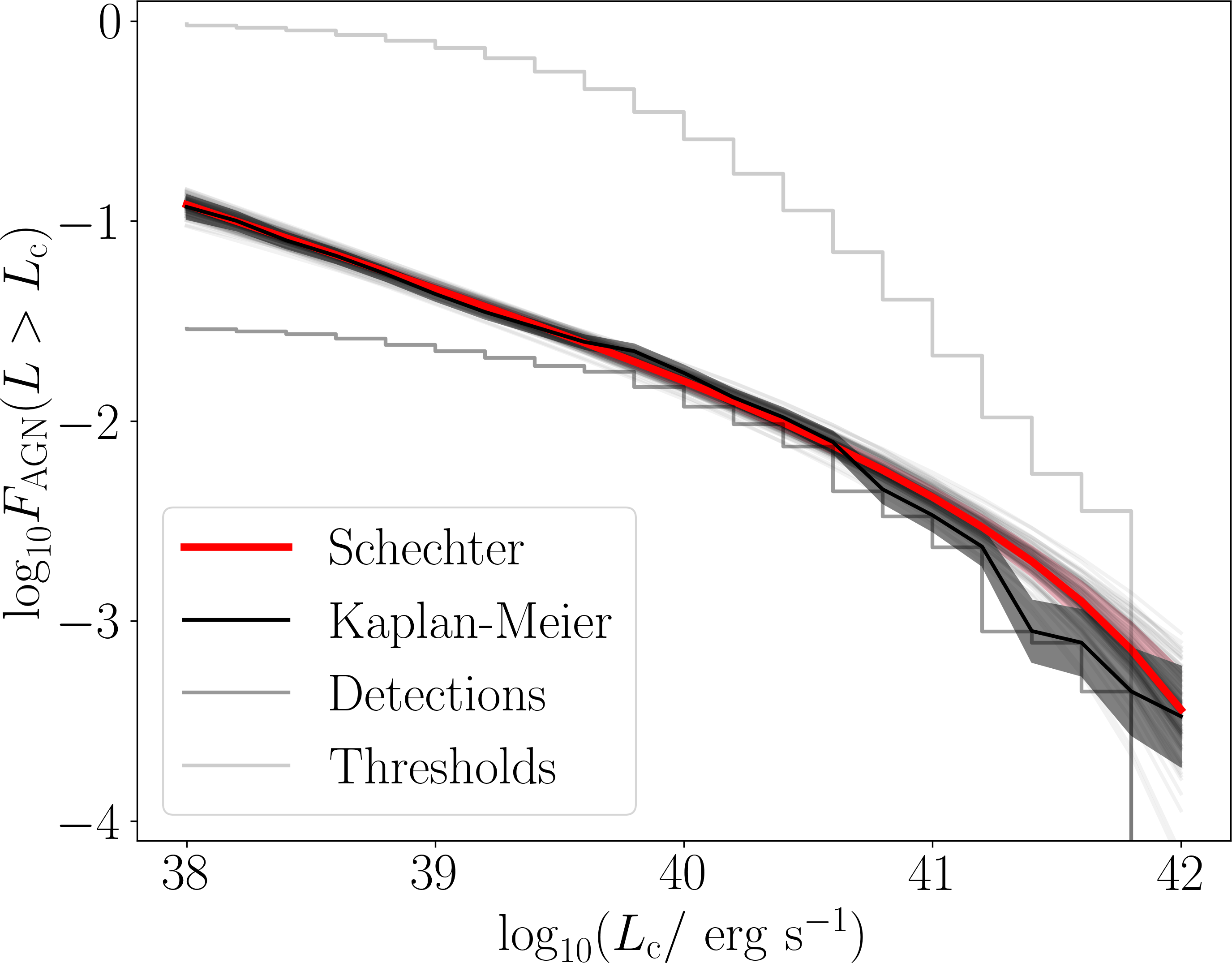}
\end{center}
\caption{\label{fig:emcee_lumhb} {\it Left panel:} Bivariate posterior
  distributions among $\alpha$, $L_\ast$, $L_{\rm min}$, and $F_{\rm
    AGN, 39}$, for the analysis of the entire sample of AGN and AGN
  nondetections, using \lhbcorr. This analysis includes all stellar
  masses and sSFRs, and we show later that the best-fit parameters
  depend strongly on those galaxy properties.  {\it Right panel:}
  $F_{\rm AGN, Y}$ as a function of $Y=\log_{10} L_c$ in erg~s$^{-1}$.
  The thick red line is the mean for the parametric Schechter function
  estimate, the thin lines are 100 samples of the posterior, and the
  pink band shows the standard deviation of $\log_{10} F_{\rm AGN}$
  around the mean. The black line is the estimate using
    \citet{kaplan58a}, with the gray band showing the $1\sigma$
    uncertainties using the formula of \citet{greenwood26a}. The
    dark-gray histogram shows the cumulative distribution of AGN
    detections and the light-gray histogram shows the cumulative
    distribution of AGN detection thresholds for nondetections (each
    normalized to the full sample of detections and nondetections). }
\end{figure}

The left panel of Figure \ref{fig:emcee_lumhb} illustrates the
bivariate posterior distributions among $\alpha$, $L_\ast$, $L_{\rm
  min}$, and $F_{\rm AGN, 39}$ for this case. The Schechter parameters
are reasonably well constrained for the full sample, but with a strong
degeneracy between $\alpha$ and $L_{\rm min}$. It is straightforward
to understand the origin of this degeneracy; for example, to increase
the number of nondetections, the model can either increase $\alpha$
or decrease $L_{\rm min}$.  Whereas these model parameters are hard to
interpret in the context of such degeneracies, $F_{\rm AGN, 39}$ is
relatively well defined, and it is does not share a degeneracy with
the model parameters (i.e. the model parameter degeneracies are along
lines of constant $F_{\rm AGN, 39}$).

The right panel of Figure \ref{fig:emcee_lumhb} shows $F_{\rm AGN, Y}$
as a function of $Y=\log_{10} L_c$ in erg~s$^{-1}$. The thick line is
the mean, the thin lines are 100 samples of the posterior, and the
pink band shows the standard deviation of $\log_{10} F_{\rm AGN}$
around the mean. The standard deviation is narrowest around $\log_{10}
F_{\rm AGN} \sim 39$, which is why we picked that threshold.

The right panel of Figure \ref{fig:emcee_lumhb} also includes the
nonparametric estimate using the approach of \citet{kaplan58a} as the
black line, with errors shown as the gray band. This estimate is in
good agreement with the Schechter function parameterization, with a
slight deviation at the high-luminosity end.

For comparison, in the right panel of Figure \ref{fig:emcee_lumhb} we
also show the raw cumulative distribution of AGN detections (dark-gray
histogram) and detection thresholds (light-gray histogram), normalized
to the sum of both populations. At the higher luminosities, the raw
cumulative distribution is representative, since those luminosities
exceed most of the detection thresholds. However, below about $L_c
\sim 3\times 10^{40}$ erg s$^{-1}$ the raw distribution starts to
undercount the true distribution owing to the selection effects. As we
show below, the luminosity at which this deviation starts to occur is
a strong function of host galaxy properties.

We can also check whether this model does a good job at explaining the
distribution of the luminosities of detected AGN. Using a
Kolmogorov-Smirnov (KS) statistic based on this distribution, we find
that it is worse than the observed sample for a fraction $p=0.43$ of
Monte Carlo samples generated under our model; that is, the observed
distribution is consistent with the expectation under the model. This
result suggests that this model is flexible enough to explain the
observations.

\subsection{Stellar mass and sSFR dependence}
\label{sec:lum_mssfr}

To study the stellar mass and sSFR dependence of the luminosity, we
have divided the sample into twelve bins, combining four in
$\log_{10}M_\ast / M_\odot$ using bin edges $[9.25, 10.25, 10.75,
  11.25, 11.75]$, with three in sSFR using bin edges $[-16.5, -13.0,
  -11.0, -9.5]$. Figure \ref{fig:mssfr} depicts these bins.

\begin{figure}[t!]
\begin{center}
  \includegraphics[width=0.98\textwidth]{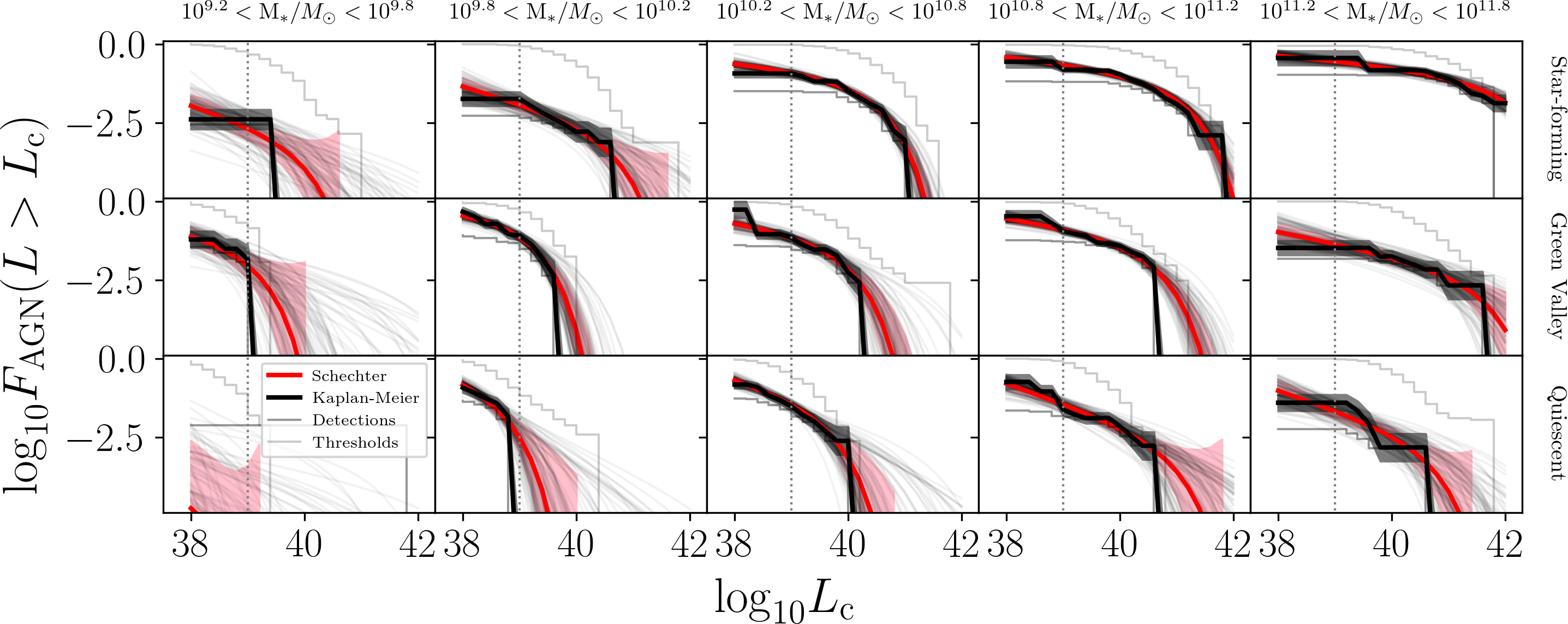}
\end{center}
\caption{\label{fig:emcee_lumhb_all} $F_{\rm AGN, Y}$ as a
  function of $Y=\log_{10} L_c$ in erg~s$^{-1}$, in bins of sSFR
  (increasing upward) and of stellar mass (increasing to the right),
  using the bins in Figure \ref{fig:mssfr}. Each panel is similar to
  the right panel of Figure \ref{fig:emcee_lumhb}, but restricted to
  galaxies in the relevant bin. The vertical dotted line shows $L_c =
  10^{39}$ erg~s$^{-1}$, the luminosity we use for defining occurrence
  rates of AGN in Figure \ref{fig:lumhb_mssfr}. }
\end{figure}

Figure \ref{fig:emcee_lumhb_all} shows for each bin the results of
performing the analysis described in Section \ref{sec:lum_uniform} to
constrain the distribution of \lhbcorr. The parametric Schechter
function estimate and the nonparametric \citet{kaplan58a} estimate
agree well in all cases, with some deviations at higher luminosities.
The individual posterior samples for each bin behave qualitatively
similarly to that for the full sample, with strong degeneracies
between $\alpha$ and $L_{\rm min}$ but with $F_{\rm AGN, 39}$ better
constrained.

\begin{figure}[t!]
\begin{center}
  \includegraphics[width=0.98\textwidth]{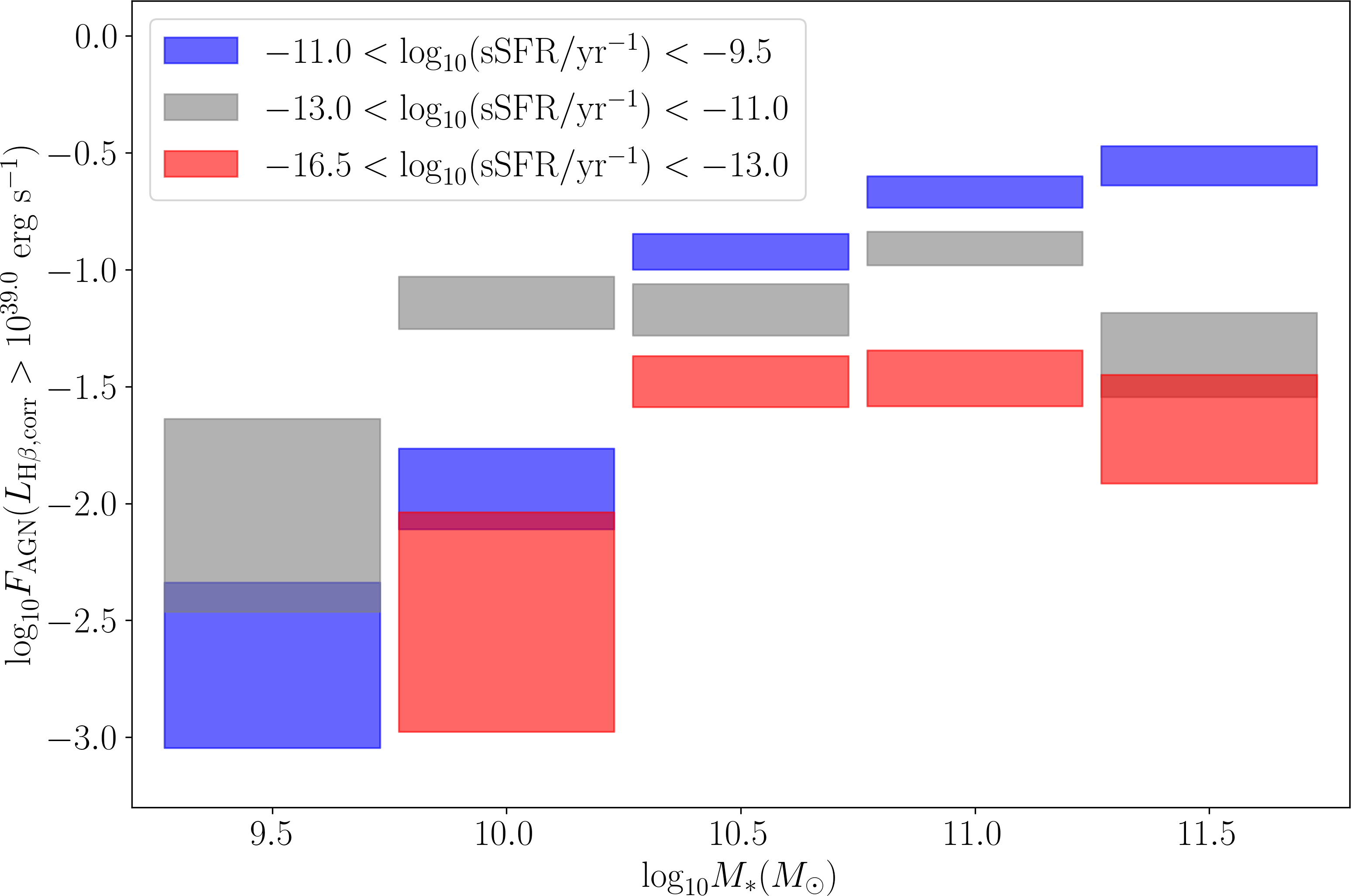}
\end{center}
\caption{\label{fig:lumhb_mssfr} AGN occurrence rate $F_{\rm
    AGN}(\lhbcorr > 10^{39}$ erg~s$^{-1})$, based on model fits
  (i.e. corrected for selection effects), as a function of stellar
  mass and sSFR. The bands are $\pm 1\sigma$ uncertainties based on
  the posterior distribution.}
\end{figure}

Figure \ref{fig:lumhb_mssfr} shows the resulting dependence of $F_{\rm
  AGN, 39}$ as a function of stellar mass and sSFR. This plot contains
the first important result of this paper: the dependence of AGN
occurrence rate as a function of galaxy properties, corrected for
selection effects, with ``AGN'' defined as having $\lhbcorr > 10^{39}$
erg~s$^{-1}$. For star-forming galaxies, there is a clear and steady
mass dependence. Quiescent galaxies have a much lower occurrence rate
of AGN and less dependence on mass. The occurrence rate for the
central sSFR bin (``green valley'' galaxies) sits in between that of
the highest and lowest sSFR bins, except for at the lowest masses
where the green valley galaxies have the highest occurrence rates.
There is also some evidence of a peak occurrence rate at $M_\ast \sim
10^{11}M_\odot$ from green valley galaxies.

There is little dependence of the results on the detection
significance we require on emission lines, if we raise the threshold
of detection from 2$\sigma$ to 3$\sigma$. If we change the luminosity
threshold on \lhbcorr\ from $10^{39}$ erg~s$^{-1}$ to $10^{40}$
erg~s$^{-1}$, $F_{\rm AGN}$ drops in all bins by about 0.2--1.0 dex,
with larger drops for the quiescent galaxies than star-forming
galaxies, but the qualitative trends remain; a difference is that for
the higher luminosity thresholds there are no mass bins in which green
valley galaxies have the highest occurrence rates.

We have also examined the luminosity distribution of $\loiiicorr$,
finding very similar trends, though with $\loiiicorr$ we find a $\sim
0.2$ dex median difference in the luminosities. There is a spread of
about 1 dex in the differences between $\loiiicorr$ and $\lhbcorr$,
but these differences only have a minor effect on the trends in the
luminosity distributions.

The thresholds depend on our choice for the nominal template AGN
(Equation \ref{eq:nominal}). We have therefore chosen an alternative
set of line ratios that are more extreme in the Seyfert direction,
equivalent to $P1 =0.544$, $P2=-0.156$, $P3=0.907$. Using these line
ratios for the nominal AGN changes the results in minor ways, similar
to or less than the uncertainties.

Finally, we have explored other choices of the $P1$--$P3$ selection
criteria. If we make them more inclusive, we do not find consistent
results---both the detection rates and the occurrence rates
increase. This result indicates that the distribution of line ratios
is not well described by the model AGN luminosity distribution beyond
these limits. If we make them less inclusive, the detection rates
decrease but the occurrence rates remain roughly constant, though the
uncertainties rise. This result suggests that the AGN model
distribution does explain the distribution of observed line ratios in
the domain of our standard selection criteria.

\section{Eddington Ratio Distributions}
\label{sec:edist}

\subsection{Methodology}
\label{sec:methodology_edr}

In this section we estimate the narrow-line Eddington ratio
distributions. For the results shown here, we estimate Eddington ratio
estimates based on the bolometric luminosity estimated from the
dust-corrected \hb\ luminosity with \citet{netzer19a} and the black
hole mass estimated from the velocity dispersion within $1R_e$ and the
$M_{\rm BH}$--$\sigma$ relation of \citet{kormendy13a} (see Section
\ref{sec:eddington_ratio}).

We use the same Schechter function form and analysis techniques that
we used for the luminosity distribution in Section
\ref{sec:methodology_lum}.

\subsection{Full sample analysis of the Eddington ratio distribution}
\label{sec:edr_uniform}

\begin{figure}[t!]
\begin{center}
  \includegraphics[width=0.48\textwidth]{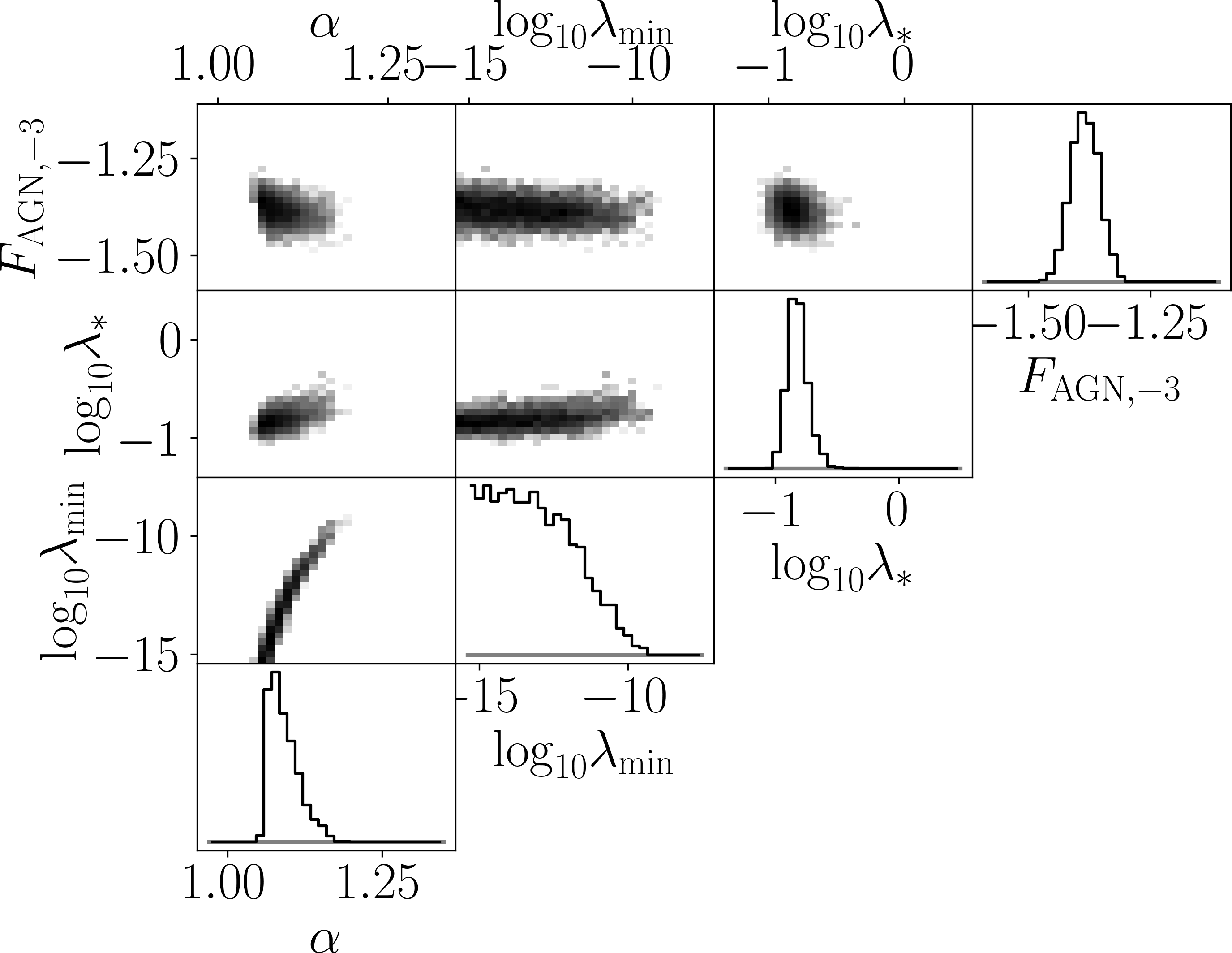}
  \quad
  \includegraphics[width=0.48\textwidth]{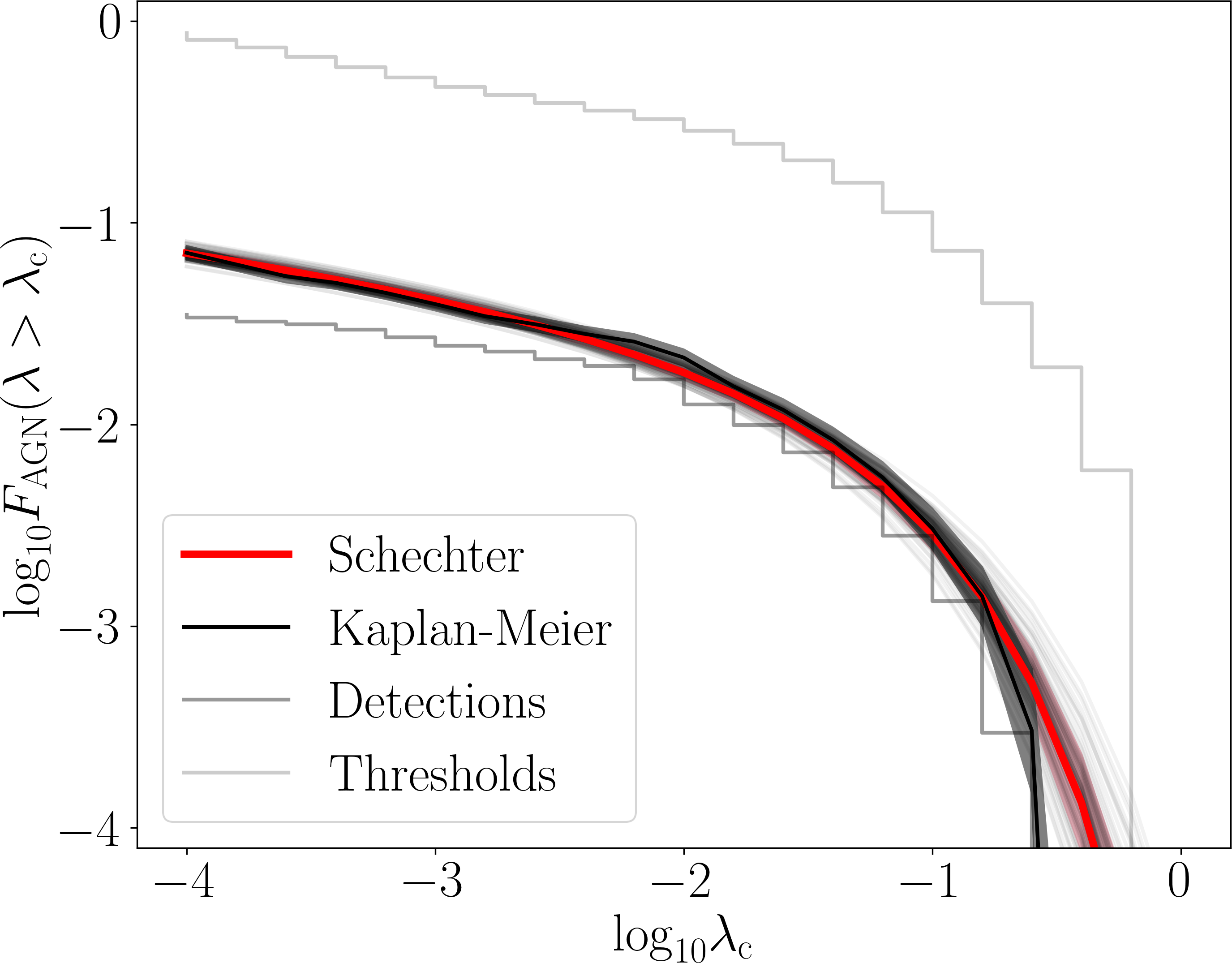}
\end{center}
\caption{\label{fig:emcee_edrhb} Similar to Figure
  \ref{fig:emcee_lumhb}, but for the Eddington ratio distribution
  instead of the luminosity distribution. The Eddington ratio is
  calculated using the calibration of \citet{netzer19a} for the
  bolometric luminosity and the $M_{\rm BH}$--$\sigma$ relationship of
  \citet{kormendy13a}. In this case, we use a reference Eddington
  ratio of $\lambda = 10^{-3}$ and examine the model-based constraints
  on $F_{\rm AGN, -3}$.} 
\end{figure}

Again as an illustration of the method, we consider the Eddington
ratio distribution from the full sample undifferentiated by stellar
mass or sSFR. Figure \ref{fig:emcee_edrhb} shows the resulting
constraints on the parameters of the distribution. In this case, we
define $F_{{\rm AGN}, -3}$ as the fraction of AGN in the model with
$\lambda > 10^{-3}$; as we found in the case of the luminosity
distribution, this quantity is relatively stable and not degenerate
with the model parameters. The Schechter fit proves a reasonably good
fit to the data in this case, with a KS statistic of $p=0.10$; the
deviation from the model driving the KS statistic low manifests as a
$\sim 30\%$ dip in the observed distribution around $\lambda_c \sim
10^{-3}$.

\subsection{Stellar mass and sSFR dependence}
\label{sec:edr_mssfr}

To investigate the stellar mass and sSFR dependence of the Eddington
ratio distribution, we use the same stellar mass and sSFR bins as we
did for the luminosity distribution in Section \ref{sec:lum_mssfr}, as
illustrated in Figure \ref{fig:mssfr}.

\begin{figure}[t!]
\begin{center}
  \includegraphics[width=0.98\textwidth]{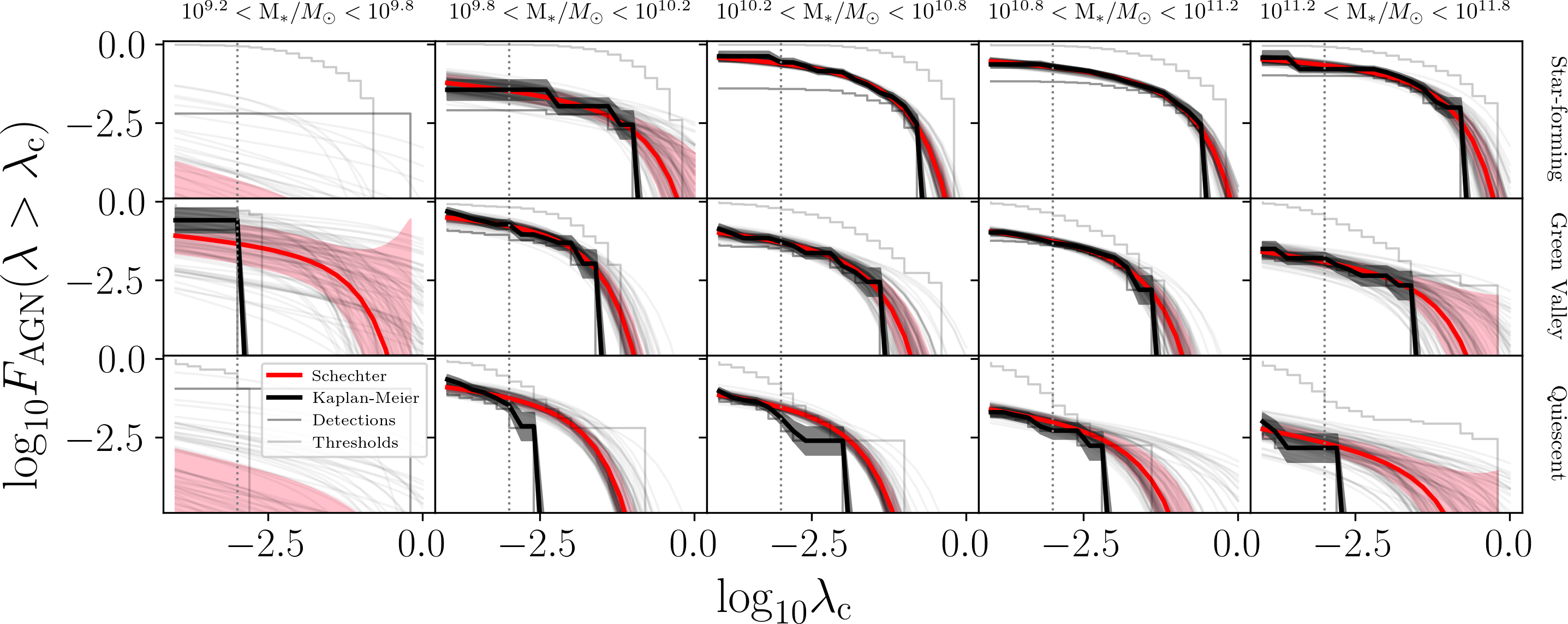}
\end{center}
\caption{\label{fig:emcee_edrhb_all} Similar to Figure
  \ref{fig:emcee_lumhb_all}, but for the Eddington ratio distribution
  instead of the luminosity function.  The Eddington ratio is
  calculated using the calibration of \citet{netzer19a} for the
  bolometric luminosity and the $M_{\rm BH}$--$\sigma$ relationship of
  \citet{kormendy13a}. The vertical dotted line shows $\lambda =
  10^{-3}$, the Eddington ratio we use for defining occurrence rates
  of AGN in Figure \ref{fig:edrhb_mssfr}. }
  
\end{figure}

Figure \ref{fig:emcee_edrhb_all} shows for each bin the constraints on
the distribution of the Eddington ratio $\lambda$. The results are
similar to those shown in Figure \ref{fig:emcee_lumhb_all}. The
parametric Schechter function estimate and the nonparametric
\citet{kaplan58a} estimate agree well, again with a handful of
deviations at higher luminosities, and the quantity $F_{\rm AGN, -3}$
is well constrained and in agreement between both methodologies.

\begin{figure}[t!]
\begin{center}
  \includegraphics[width=0.98\textwidth]{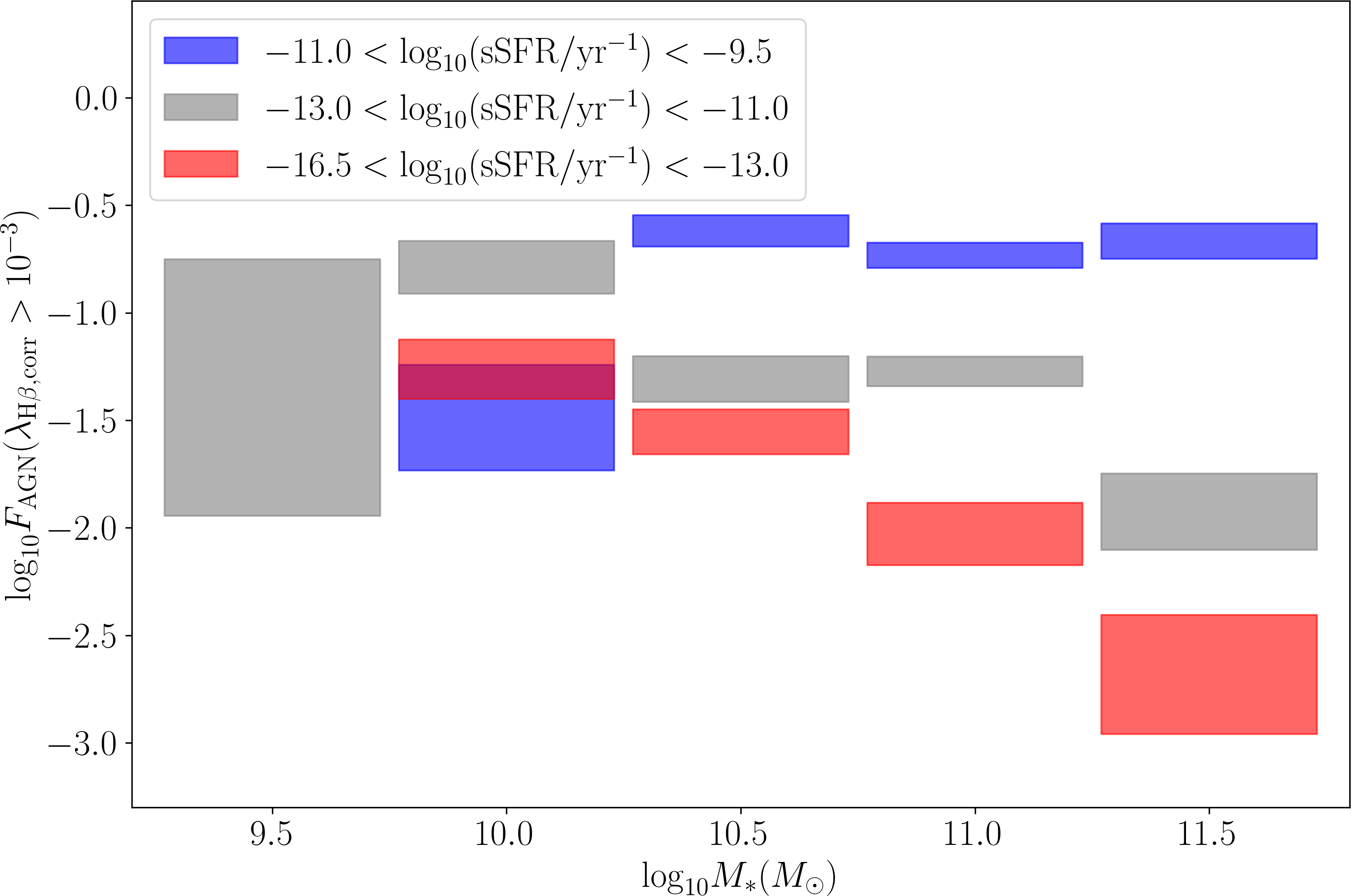}
\end{center}
\caption{\label{fig:edrhb_mssfr} AGN occurrence rate $F_{\rm
    AGN}(\lambda > 10^{-3})$, based on model fits (i.e. corrected for
  selection effects), as a function of stellar mass and sSFR. The
  bands are $\pm 1\sigma$ uncertainties based on the posterior
  distribution. For this plot we use the $M$--$\sigma$ relation of
  \citet{kormendy13a} and bolometric luminosities based on
  dust-corrected H$\beta$ and the conversion of \citet{netzer19a}.}
\end{figure}

Figure \ref{fig:edrhb_mssfr} shows the results for $F_{{\rm AGN}, -3}$
as a function of stellar mass and sSFR. For this figure we use
bolometric luminosities based on the dust-corrected \lhbcorr, and the
relation of \citet{netzer19a} (Equation \ref{eq:netzer}), and the
$M_{\rm BH}$--$\sigma$ relation of \citet{kormendy13a}. The results
are close to as expected based on Figure \ref{fig:lumhb_mssfr}. Using
a Eddington ratio threshold instead of a luminosity threshold is
similar to (though not exactly the same as!) dividing by stellar mass.
Therefore it tilts the stellar mass dependence downward. The result
is that at the higher masses the dependence on stellar mass is flat
for star-forming galaxies, declining strongly for the quiescent
galaxies, and somewhere in between for the green valley galaxies. The
lowest mass bin is again somewhat anomalous---for galaxies with $9.25
< \log_{10} M_\ast / M_\odot <10.25$, there is a weak detection that
green valley galaxies have the highest AGN occurrence rates.

We have explored the use of other $M_{\rm BH}$--$\sigma$ calibrations
and bolometric luminosity determinations. The $M_{\rm BH}$--$\sigma$
from \citet{kormendy13a} yields somewhat lower $F_{{\rm AGN}, -3}$, by
about 0.1--0.3 dex depending on the bin, relative to the calibrations
of \citet{greene06a}, \citet{gultekin09a}, or \citet{graham13a}, which
all yield results similar to each other. Otherwise, the results are
qualitatively similar. The offset for \citet{kormendy13a} arises
because their value of $\alpha$ (in Equation \ref{eq:msigma}) is high,
leading to higher black hole masses and consequently lower Eddington
ratios.

The bolometric luminosities based on dust-corrected \lhbcorr\ 
(\citealt{netzer19a}; Equation \ref{eq:netzer}) yield results very
similar to those based on the dust-corrected \loiiicorr\ 
(\citealt{kauffmann09a}; Equation \ref{eq:kauffmann}). The bolometric
luminosities based on the uncorrected \loiii\ (\citealt{heckman04a};
Equation \ref{eq:heckman}) are fairly similar but yield somewhat
weaker dependence of $F_{{\rm AGN}, -3}$ on sSFR. This difference may
be related to higher dust corrections in high-sSFR galaxies, which in
the case of the dust-corrected luminosities widens the difference in
bolometric luminosities in these classes of galaxy, though the
difference in the median dust corrections is only of order 0.2 dex
between the low- and high-sSFR AGN samples.

As for the luminosity dependence of $F_{\rm AGN}$, there is little
dependence on the detection significance (raising the threshold of
line detection from 2$\sigma$ to 3$\sigma$). If we change the
threshold $\lambda_c$ from $10^{-3}$ to $10^{-2}$, $F_{\rm AGN}$ drops
in all bins by about 0.2--0.8 dex, again with larger drops for the
quiescent galaxies than star-forming galaxies. If we change the choice
of the nominal AGN line ratios, the results alter only
slightly. Otherwise, the results are qualitatively similar.

\section{Comparison with Previous Results}
\label{sec:comparison}

Our analysis can be directly compared to the one other study that we
have found of narrow-line AGN dependence on stellar mass and sSFR that
accounts for selection effects in assessing their demographics, that
of \citet{trump15a}. They study the SDSS Legacy Survey Main Sample
galaxies (\citealt{york00a, strauss02a}), which is a far larger sample
than MaNGA, though with lower signal-to-noise ratio spectra and poorer
measurements of stellar mass and sSFR. Using a narrow-line selection
criterion based on \citet{baldwin81a}, their Figure 13 shows the
detected AGN fraction as a function of stellar mass and sSFR, compared
to a best-fit model with a uniform Eddington ratio distribution. The
residuals are somewhat similar to our results but not exactly the
same: relative to a constant Eddington ratio distribution, AGN
occurrence increases with stellar mass for star-forming galaxies (more
noticeably than we find), and decreases with stellar mass for
quiescent galaxies. The nature of the analysis in \citet{trump15a}
does not make a direct quantitative comparison easy to make.

Our results are not appropriate for comparison with most other
investigations of the stellar mass and sSFR dependence of narrow-line
AGN occurrence, because all other such investigations we found in the
literature do not account for the selection effects. MaNGA has the
highest signal-to-noise ratios available for a statistically
interesting sample, i.e. considerably higher than the SDSS Legacy or
DESI samples for typical galaxies. The MaNGA sample is also at lower
redshifts, allowing the central spectrum to be better isolated from
star-formation-related contamination. Yet even for MaNGA we have shown
that the detected AGN demographics are primarily shaped by the
selection effects. Any analysis of narrow-line AGN host galaxy
demographics that omits a consideration of selection effects is
therefore likely to be inaccurate except for the most luminous AGN
(e.g. \citealt{kauffmann03a, heckman04a, martin07a, salim07a,
  kauffmann09a, schawinski10a, ellison16a, sanchez18a, man19a,
  greene20b, mcdonald21a, woodrum24a, pucha25a}). Many of these
studies conclude that narrow-line AGN are most frequent for galaxies
with intermediate sSFRs, i.e. in the green valley, without
investigating whether this effect is caused by the difficulty of
detecting narrow-line AGN in galaxies with high sSFR.

It is worth comparing to the Eddington ratio distributions published
by \citet{kauffmann09a} as a function of star formation history. In
that paper, they use $\loiii / M_\ast({\rm BH})$, proportional to the
Eddington ratio. The quiescent galaxies show a power-law distribution
of this quantity in their analysis, with a dependence on mass similar
to ours. In their case, they mix LINERS and Seyferts, so at low
measured Eddington ratio, it is not obvious that they are measuring
AGN, but at high Eddington ratios they are likely to be. In contrast,
star forming galaxies show a lognormal peak in the Eddington ratio
distribution in their analysis, and they describe this situation as
the ``feast mode,'' with a universal lognormal Eddington ratio
distribution.

\citet{jones16a} convincingly showed that the shape of the ``feast''
mode in \citet{kauffmann09a} is nearly entirely shaped by selection
effects, and the intrinsic distribution can be well described by a
Schechter function.  Our analysis here supports that
interpretation. Quantitatively, however, their determined faint-end
slope $\alpha \sim 1.38$\footnote{\citet{jones16a} report
$\alpha \sim 0.38$ but their Equation (1) shows that the $\alpha$ they
are reporting is the log-slope of $\Phi(\log_{10} \lambda)$, and
therefore our $\alpha$ must be compared to their $\alpha+1$.}  is
steeper than our values for the full sample of $1.1$ or so, and they
determine $\log_{10} \lambda_{\rm min} = -3.9$, which means that
essentially every galaxy has an AGN with at least that Eddington
ratio, whereas for our full sample we infer that less than 10\% of
galaxies are above this minimum. Their sample is distributed
differently in stellar mass than MaNGA, which is flat between $10^9$
and $10^{10}$ $M_\odot$, but even for our most massive, star-forming
sample we find the fraction of galaxies with $\log_{10} \lambda >
-3.9$ to be at most 50\%.

The most similar analyses to those we present here have been performed
using X-ray data, in particular by \citet{aird12a} and subsequent work
from the same group (\citealt{aird19a, birchall23a}), which accounts
for the selection effects and infers $F_{\rm AGN}$ as a function of
mass and (for one narrow range of masses) SFR. In comparing the
results, we note that these investigators assume $M_{\rm BH} \sim
0.002 M_{\rm \ast}$, whereas for the MaNGA stellar mass estimates and
the $M_{\rm BH}$--$\sigma$ relation of \citet{kormendy13a}, that ratio
only holds at the highest masses ($\log_{10} M_\ast \sim 11.5$) and is
much smaller at lower mass, by about 1.0 dex at $\log_{10} M_\ast \sim
10$.  These lower black hole mass estimates will cause our Eddington
ratio estimates to be larger for many galaxies. Assuming that the
bolometric corrections are correct, this choice makes their $\log_{10}
\lambda_c$ choice of $-3.5$ in their analysis a bit more comparable to
our choice of $-3$ on average.

Although our measurements are qualitatively similar to those of
\citet{birchall23a}, whose analysis is at redshifts comparable to
ours, quantitatively we find a substantial difference. For
star-forming galaxies, we find a larger fraction of narrow-line AGN
than they do of X-ray AGN, of nominally comparable Eddington ratios,
by about a factor of ten. Meanwhile, for quiescent galaxies relative
to star-forming galaxies, \citet{birchall23a} find a smaller fraction
of X-ray AGN than we do, but by a much smaller difference than we do
(a roughly constant factor of 2--4, rather than a mass-dependent
factor reaching about 100). The low-redshift X-ray results in the work
of \citet{aird19a} are similar to \citet{birchall23a}, and the results
of \citet{wang17a} and \citet{torbaniuk21a} qualitatively agree.

The meaning of these disagreements between our narrow-line analysis
and the X-ray analyses is unclear. They could reflect errors in our
bolometric corrections for narrow-line regions. For example, for
massive quiescent galaxies, a deficit of warm ionized gas may lead to
smaller narrow-line luminosities at a given bolometric luminosity. On
the other hand, the disagreements could be due to our use of $M_{\rm
  BH}$-$\sigma_e$ rather than an $M_{\rm BH}/M_\ast$ ratio like
\citet{aird19a} and \citet{birchall23a}; our choice would lead to a
relatively stronger decline of $F_{\rm AGN}$ with mass, as seen, but
whether this difference explains the disagreement and the fact that it
is primarily for quiescent galaxies is unclear. This disagreement is a
problem best addressed with more direct comparisons (that still
properly account for selection effects in both sets of measurements).

\section{Summary and Discussion}
\label{sec:conclusion}

We have estimated AGN narrow-line luminosities and luminosity
detection thresholds for the MaNGA sample, as well as the Eddington
ratios and Eddington ratio detection thresholds. The detection
thresholds in our sample are important---for most AGN nondetections
these thresholds are similar to the luminosities found for the AGN
detections. Clearly many AGN are hiding in the sample that cannot be
detected, primarily due to contamination by star-formation-related
line emission. By including the thresholds in our analysis, we
estimate the actual AGN occurrence rates $F_{\rm AGN}$, meaning for
this purpose the fraction of galaxies with AGN above some fixed
luminosity or Eddington ratio.

We find a fairly complicated dependence of $F_{\rm AGN}$ on stellar
mass and sSFR, with the fraction of galaxies with Eddington ratios
$\lambda>10^{-3}$ ($F_{\rm AGN, -3}$) constant or increasing with
stellar mass for star-forming galaxies and decreasing with stellar
mass for quiescent galaxies. At higher masses ($M>10^{10.25} M_\odot$)
the occurrence rate $F_{\rm AGN}$ increases with sSFR, whereas in our
lowest mass bin $F_{\rm AGN}$ peaks for green valley galaxies (at low
significance).

Comparing with the X-ray analyses of \citet{birchall23a}, for
star-forming galaxies we find a larger AGN fraction at comparable
Eddington ratio criteria, and for quiescent galaxies we find a much
more mass-dependent AGN fraction with a much larger difference from
the star-forming galaxies at high mass. Understanding the nature of
these differences requires more extensive investigation.

Over the next few years, DESI will release a public catalog of
moderate-resolution spectra of low redshift galaxies that dwarfs the
SDSS Legacy sample and the MaNGA sample. However, this catalog will
only be usable for statistical studies of the host galaxy demographics
of narrow-line AGN if its selection effects are
characterized. \citet{trump15a} presented a method to determine
selection thresholds. Here we have extended this method to more
cleanly use nondetections of individual lines and have shown how to
use the thresholds directly in Bayesian or maximum-likelihood fits
to luminosity or Eddington ratio distributions. Although we have
applied this method to the narrow-line AGN classification method of
\citet{ji20a}, it is applicable to the more traditional classification
methods as well.

There nevertheless are still a number of flaws in our analysis that
could be better tested or addressed. First, any analysis that depends
on the bolometric luminosity, as our investigation of the Eddington
ratio distribution does, depends on accurate bolometric
corrections. Particularly in the realm of low-mass galaxies (and thus
intermediate-mass black holes), it is possible that \oiiihb\ depends
on luminosity owing to the expected change in hardness of the ionizing
spectrum (e.g. \citealt{cann19a}), which would require us to adjust
our estimates of the selection effects. Similar concerns surround the
dust corrections.

Second, our analysis has excluded the broad-line AGN, many of which
have narrow-line emission as well. This effect would be minor for our
analysis if the obscuration fraction were independent of stellar mass
and sSFR; it would just mean that we were missing a relatively small
fraction of narrow-line AGN. But we have not shown that this is the
case.

Third, our detection threshold analysis assumes that all AGN would
have the same set of strong line ratios if they were observed
uncontaminated. This uniformity is certainly not the case, and we have
not ruled out that our inferred trends are sensitive to this
assumption. As a related point, our method does not model the
distribution of non-AGN line ratios and relies on sharp detection
criteria for AGN; therefore, the results are still somewhat sensitive
to these criteria (Section \ref{sec:lum_mssfr}).

In short, although our analysis has advanced the state of the art in
accounting for selection criteria in studying AGN demographics, there
are many new samples to investigate and many improvements to our
techniques to consider. We encourage future investigators to improve
upon these methods and to apply them to the massive new samples
becoming available.

\begin{acknowledgements}
  We thank Renbin Yan and Xihan Ji for useful discussions about
  narrow-line AGN selection methods, Julia Comerford for useful 
  discussions about selecting AGN in MaNGA, and Shobita Satyapal for pointing
  us to the work on the potential dependence of \oiiihb\ on mass.
  We thank Janelle Sy for her contributions to an early investigation 
  of this dataset. An anonymous referee made many thoughtful
  comments and suggestions which substantially improved this
  manuscript, for which we thank them.
  
Funding for the Sloan Digital Sky 
Survey IV has been provided by the 
Alfred P. Sloan Foundation, the U.S. 
Department of Energy Office of 
Science, and the Participating 
Institutions. 

SDSS-IV acknowledges support and 
resources from the Center for High 
Performance Computing  at the 
University of Utah. The SDSS 
website is www.sdss4.org.

SDSS-IV is managed by the 
Astrophysical Research Consortium 
for the Participating Institutions 
of the SDSS Collaboration including 
the Brazilian Participation Group, 
the Carnegie Institution for Science, 
Carnegie Mellon University, Center for 
Astrophysics | Harvard \& 
Smithsonian, the Chilean Participation 
Group, the French Participation Group, 
Instituto de Astrof\'isica de 
Canarias, The Johns Hopkins 
University, Kavli Institute for the 
Physics and Mathematics of the 
Universe (IPMU) / University of 
Tokyo, the Korean Participation Group, 
Lawrence Berkeley National Laboratory, 
Leibniz Institut f\"ur Astrophysik 
Potsdam (AIP),  Max-Planck-Institut 
f\"ur Astronomie (MPIA Heidelberg), 
Max-Planck-Institut f\"ur 
Astrophysik (MPA Garching), 
Max-Planck-Institut f\"ur 
Extraterrestrische Physik (MPE), 
National Astronomical Observatories of 
China, New Mexico State University, 
New York University, University of 
Notre Dame, Observat\'ario 
Nacional / MCTI, The Ohio State 
University, Pennsylvania State 
University, Shanghai 
Astronomical Observatory, United 
Kingdom Participation Group, 
Universidad Nacional Aut\'onoma 
de M\'exico, University of Arizona, 
University of Colorado Boulder, 
University of Oxford, University of 
Portsmouth, University of Utah, 
University of Virginia, University 
of Washington, University of 
Wisconsin, Vanderbilt University, 
and Yale University.
\end{acknowledgements}

\appendix

\section{Analysis of Quiescent and Star-forming Galaxy Hosts}

\begin{figure}[t!]
\begin{center}
  \includegraphics[width=0.48\textwidth]{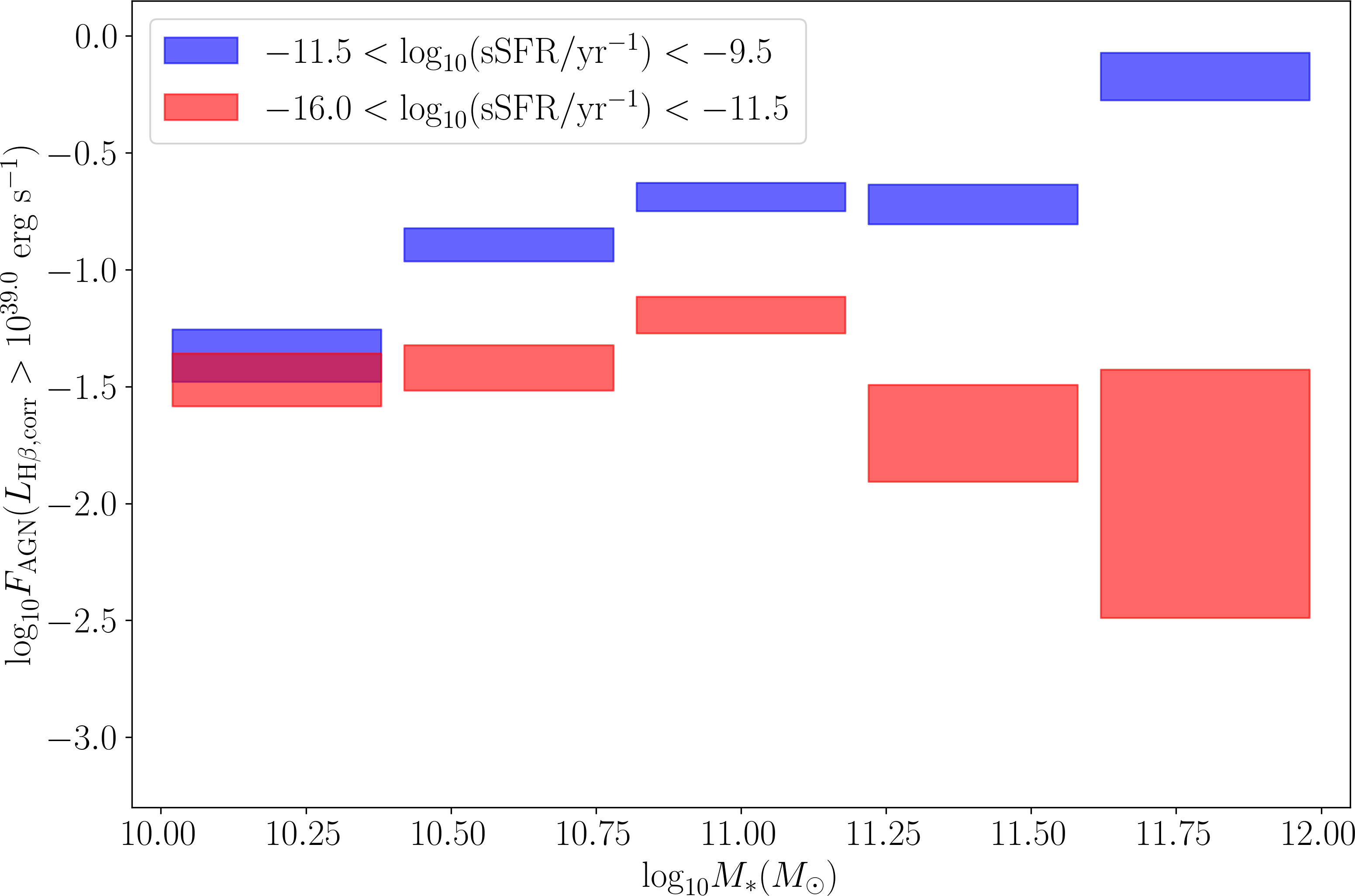}
  \includegraphics[width=0.48\textwidth]{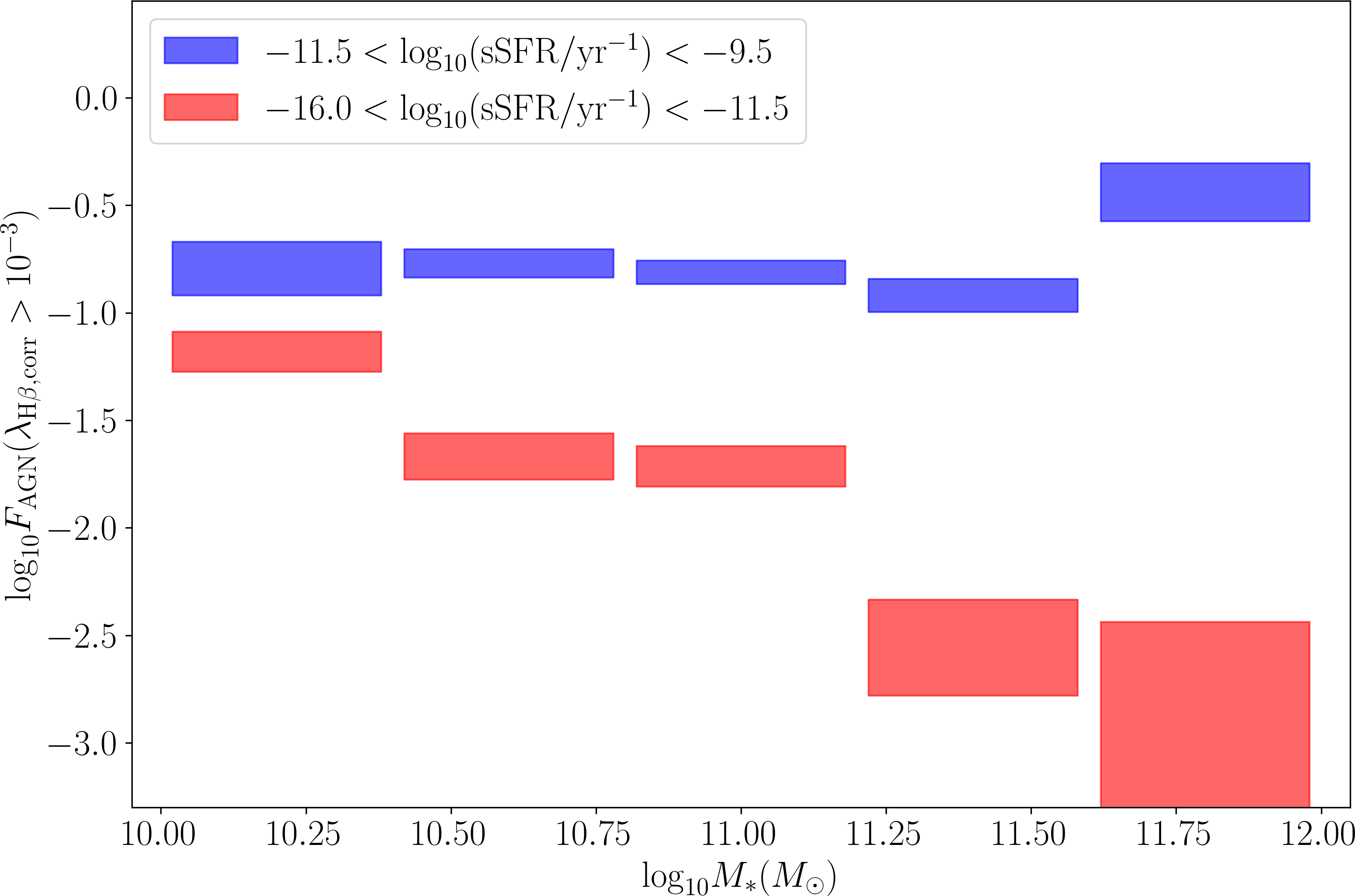}
\end{center}
\caption{\label{fig:lumhb_mssfr_arjun} Left panel: Similar to Figure
  \ref{fig:lumhb_mssfr}, with a different division of samples
  according to sSFR. Right panel: Similar to Figure
  \ref{fig:edrhb_mssfr}, with a different division of samples
  according to sSFR.}
\end{figure}

We have performed the same analysis shown in Figure
\ref{fig:lumhb_mssfr} for different choices of the sSFR bins. In
particular, for the purposes of a companion effort comparing our
results to numerical simulations of galaxy formation, we have divided
the sSFRs into two bins, one star-forming and one quiescent,
corresponding to $\log_{10}$ sSFR greater or less than $-11.5$.
Figure \ref{fig:lumhb_mssfr_arjun} shows $F_{\rm AGN}$ as a function
of stellar mass for each category. These results show the same general
trends as Figures \ref{fig:lumhb_mssfr} and \ref{fig:edrhb_mssfr}. It
is worth noting that the hint of increase in $F_{\rm AGN}$ as a
function of $\lambda$ for star-forming galaxies in Figure
\ref{fig:edrhb_mssfr}, based on the lowest mass bin, disappears in
this rebinning.

\section{Variation with Methodology }

\begin{figure}[t!]
\begin{center}
  \includegraphics[width=0.48\textwidth]{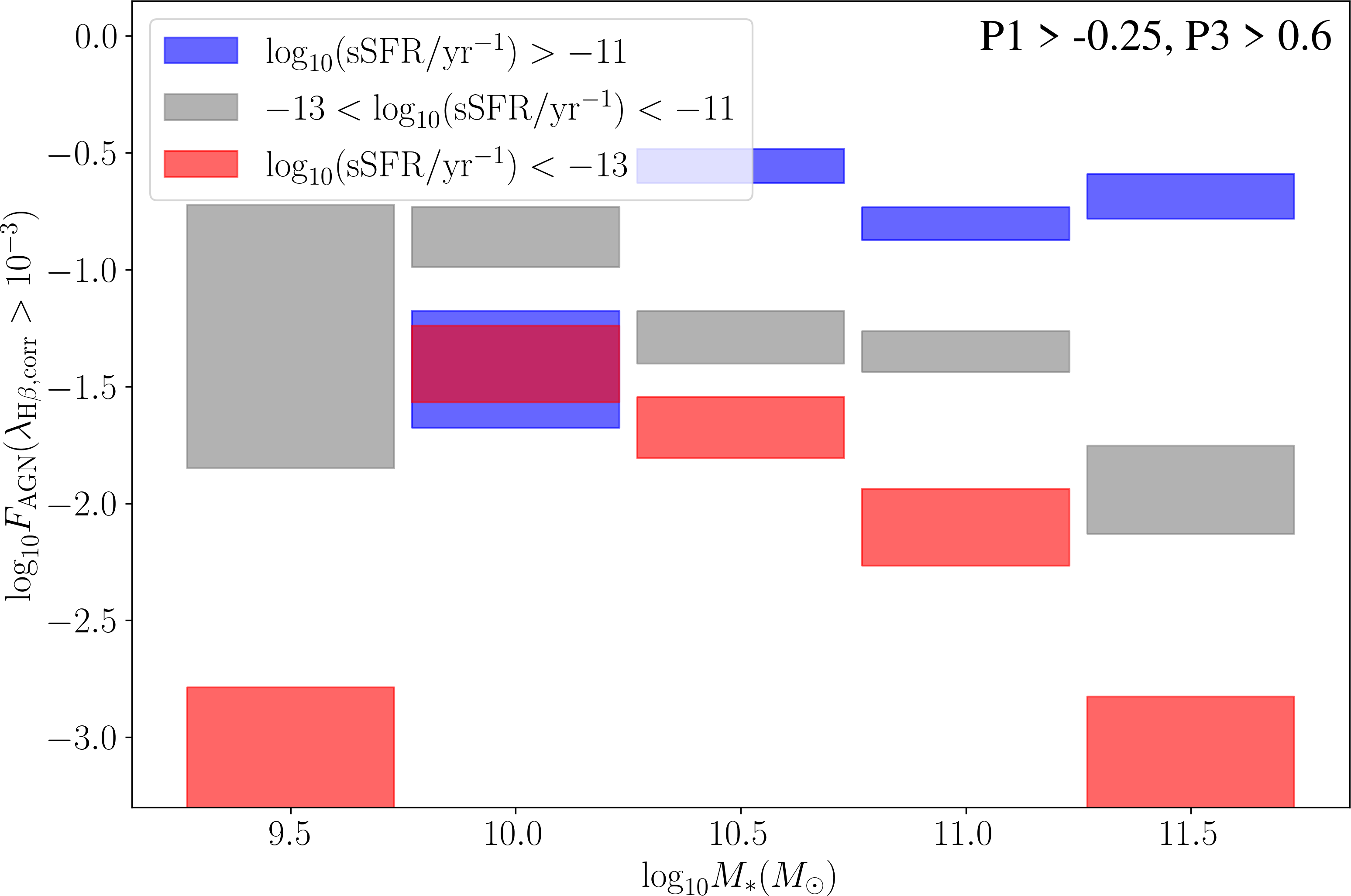}
  \includegraphics[width=0.48\textwidth]{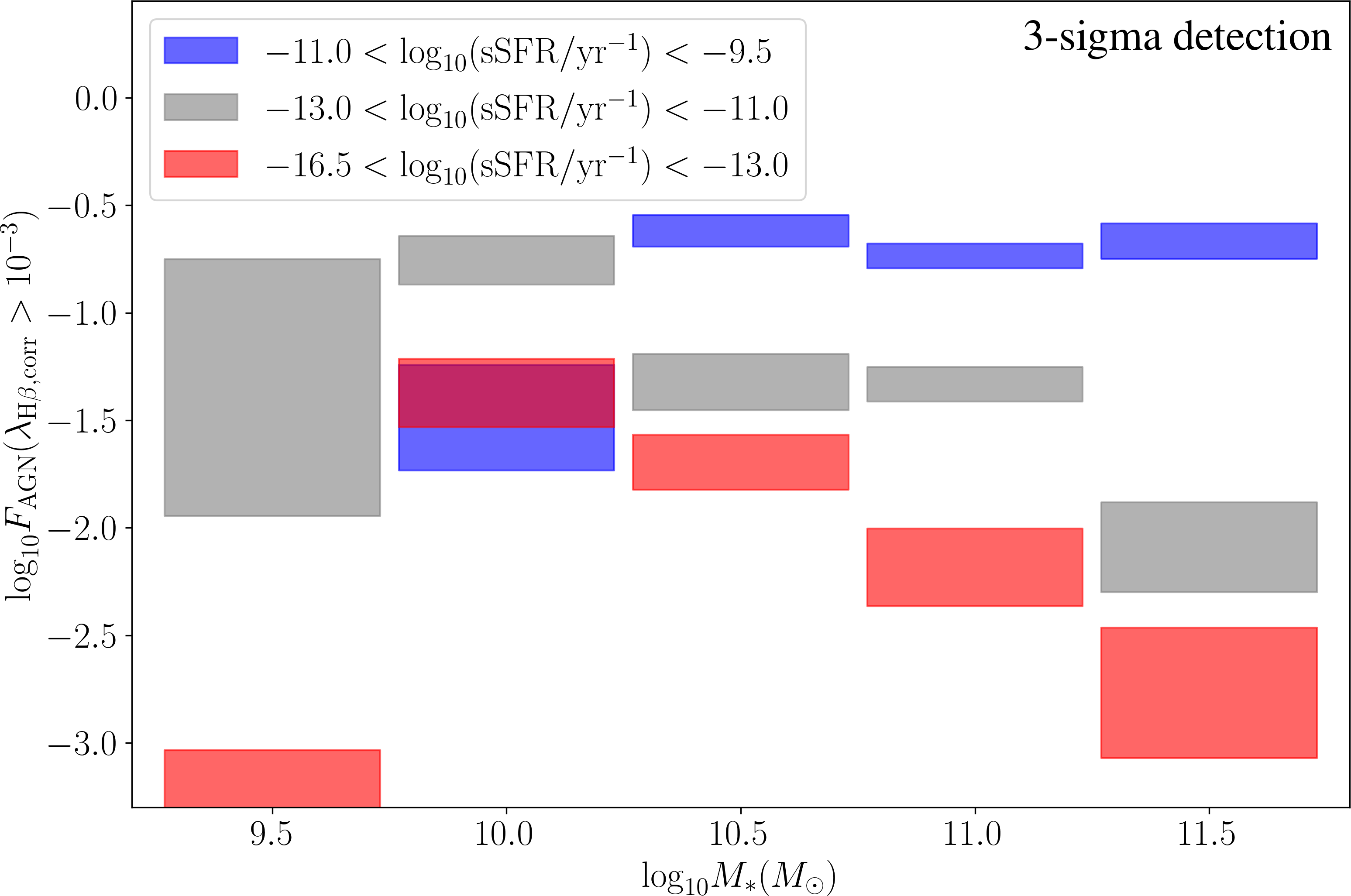}
  \\
  ~\\
  \includegraphics[width=0.48\textwidth]{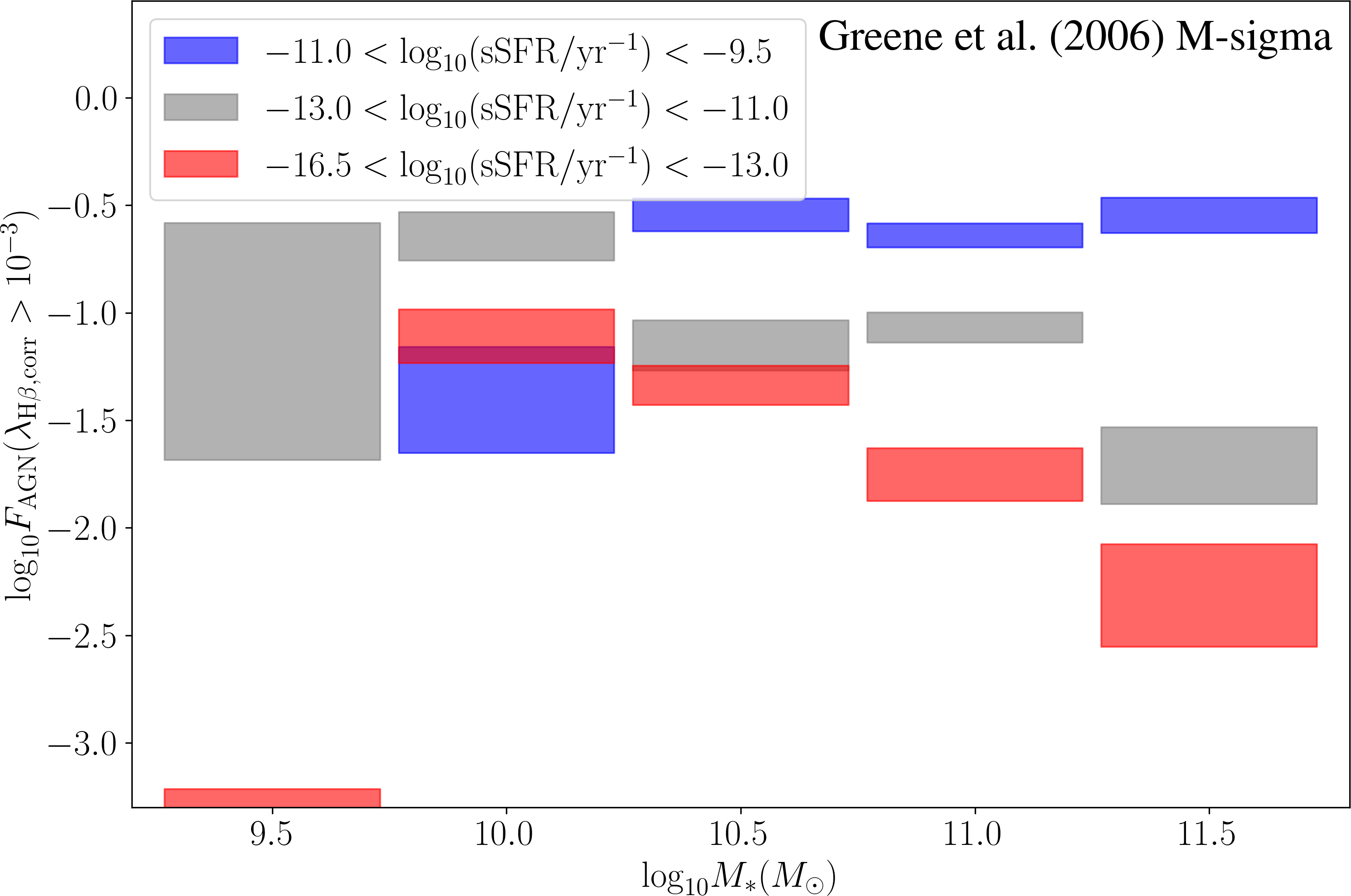}
  \includegraphics[width=0.48\textwidth]{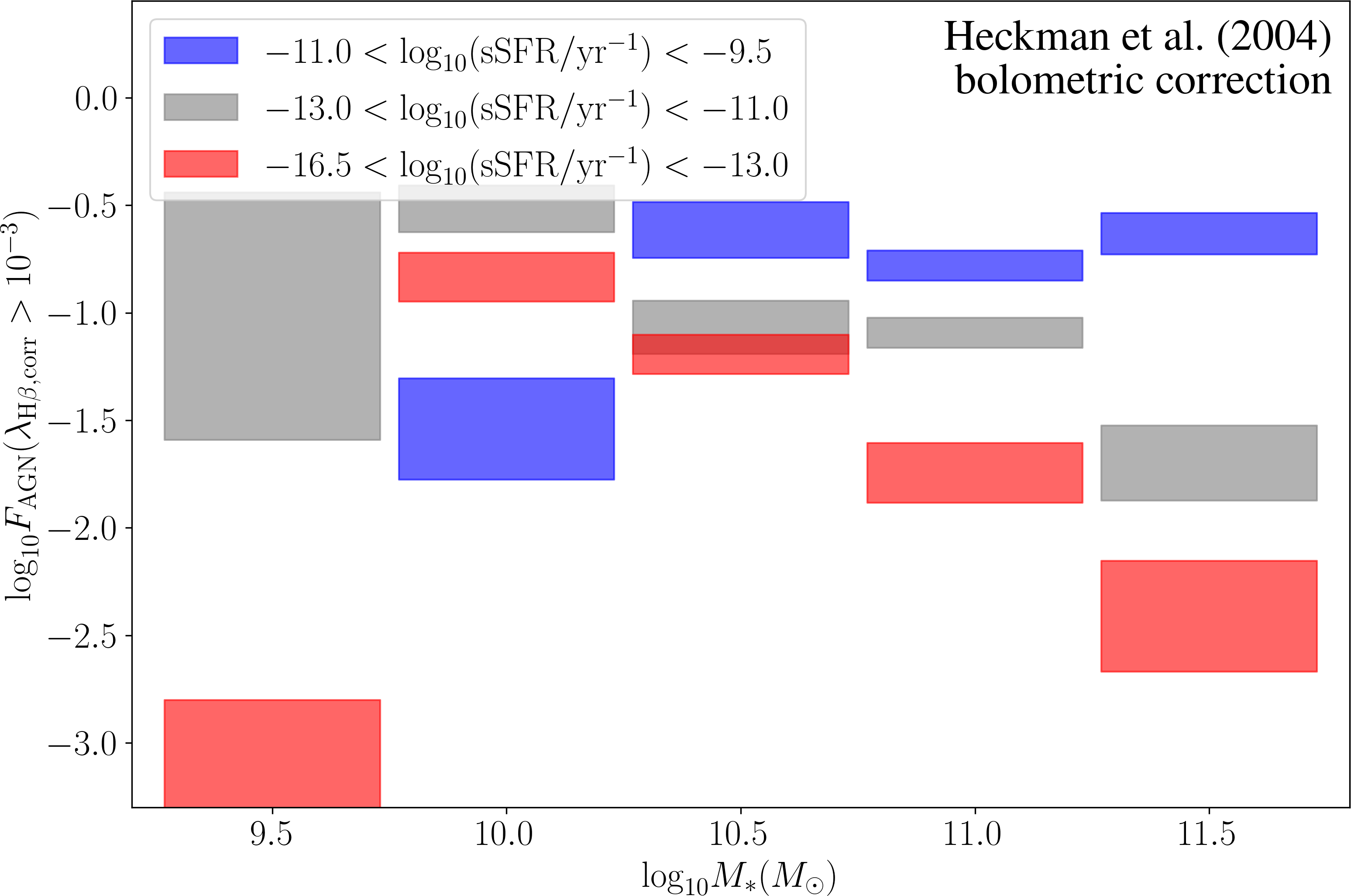}
\end{center}
\caption{\label{fig:methods} Each panel is similar to Figure
  \ref{fig:edrhb_mssfr}, with different methodological choices
  relative to our nominal choices. In Figure \ref{fig:edrhb_mssfr}, we
  use selection criteria that require 2$\sigma$ detections of each
  line and $P1 > -0.3$ and $P3 > 0.55$ to define a Seyfert, we use
  \citet{kormendy13a} to estimate the black hole mass, and we use the
  dust-corrected \lhbcorr and the relationship in \citet{netzer19a} to
  estimate the bolometric luminosity.}
\end{figure}

In Figure \ref{fig:methods}, we show how variations in methodology
affect our calculation of $F_{\rm AGN}(\lambda > 10^{-3})$. In Section
\ref{sec:agn-derived} we described various choices of the $M_{\rm
  BH}$-$\sigma_e$ relationship and the bolometric luminosity
correction. In Section \ref{sec:lum_mssfr}, we described the effects
of changing the AGN selection criteria and changing the line detection
threshold to 3$\sigma$ instead of 2$\sigma$. The figure shows
representative cases of these choices. 

In the top left panel, we show the results of making the selection
slightly more restrictive, using the criteria shown, and leaving
everything else the same.  In the top right panel, we show the effect
of requiring 3$\sigma$ detection of each line. In the bottom left
panel, we show the effect of using the \citet{greene06a} $M_{\rm
  BH}$-$\sigma_e$ relation, which is the most different from our
nominal case of \citet{kormendy13a}. In the bottom right panel, we
show the result of using the \citet{heckman04a} bolometric correction,
which is the most different from the other two we consider.

Although quantitatively differences appear, qualitatively all of these
choices yield the same basic results shown in Figure
\ref{fig:emcee_edrhb} for the nominal case. 

\bibliography{blanton}{} \bibliographystyle{aasjournal}



\end{document}

%% file: defs.tex
\newcommand{\ha}{\textrm{H}\ensuremath{\alpha}}
\newcommand{\hb}{\textrm{H}\ensuremath{\beta}}

\newcommand{\nii}{[\textrm{N}~\textsc{ii}]}
\newcommand{\sii}{[\textrm{S}~\textsc{ii}]}

\newcommand{\oiii}{[\textrm{O}~\textsc{iii}]}
\newcommand{\oiv}{[\textrm{O}~\textsc{iv}]}

\newcommand{\oiiilam}{[\textrm{O}~\textsc{iii}]~\ensuremath{\lambda5007}}
\newcommand{\niilam}{[\textrm{N}~\textsc{ii}]~\ensuremath{\lambda6584}}

\newcommand{\siidoublet}{[\textrm{S}~\textsc{ii}]~\ensuremath{\lambda\lambda6716,6731}}

\newcommand{\niiha}{\nii/\ha}
\newcommand{\siiha}{\sii/\ha}
\newcommand{\oiiihb}{\mbox{[\textrm{O}~\textsc{iii}]/\textrm{H}\ensuremath{\beta}}}

\newcommand{\loiii}{\mbox{{\ensuremath{L}([\textrm{O}~\textsc{iii}])}}}
\newcommand{\llimoiii}{\mbox{{\ensuremath{L_{\textrm{lim}}}([\textrm{O}~\textsc{iii}])}}}
\newcommand{\llimhb}{\mbox{{\ensuremath{L_{\textrm{lim}}}(\textrm{H}\ensuremath{\beta})}}}

\newcommand{\lhb}{\ensuremath{L(\textrm{H}\beta)}}

\newcommand{\lhbcorr}{\ensuremath{L(\textrm{H}\beta)_{\rm corr}}}

\newcommand{\loiiicorr}{\loiii\ensuremath{_{\rm corr}}}